\documentclass[epj]{svjour}
\pdfoutput=1
\usepackage{latexsym,wasysym,amsfonts,amssymb}
\usepackage{color}
\usepackage{graphics}
\graphicspath{{figspaper/}}
\definecolor{mygreen}{rgb}{0,0.5,0}

\def\beq {\begin{equation}}
\def\eeq {\end{equation}}
\def\beqa {\begin{eqnarray}}
\def\eeqa {\end{eqnarray}}
\def \bnum {\begin{enumerate}}
\def \enum {\end{enumerate}}
\def\bi {\begin{itemize}}
\def\ei {\end{itemize}}

\def\th{\theta}
\def\omg{\Omega}
\def\omgv{\mathbf{\Omega}}

\def\tomg{\tau_{\Omega}}
\def\tnr{t/T_{E}^{0}}
\def\tinit{T_{E}^{0}}

\def\tomgnr{{t/\tomg}}

\def\ro{{Ro}_T}
\def\komg{k_\Omega}
\def\komgnr{k_{\Omega} \eta}
\def\knorm{k/k_\Omega}

\def\uu { \mathbf{u} }
\def\ux { \mathbf{x} }
\def\unab { \mathbf{\nabla} }
\def\dho{\partial}
\def\ur { \mathbf{r} }
\def\mbf {\mathbf}
\def\kmax {k_{max}}
\def\delv { \mathbf{\Delta} }
\def\la{\langle}
\def\ra{\rangle}

\def\ss{\scriptstyle}
\def\bla {\Big{\langle}}
\def\bra {\Big{\rangle}}
\def\kd{k^{'}}
\def\kdv{\mathbf{k^{'}}}
\def\kv{\mathbf{k}}
\def\uihat{\hat{u}_i}
\def\ujhat{\hat{u}_j}
\def\dl{\displaylimits}

\begin{document}
%
\title{Rotating turbulence under ``precession-like" perturbation 
}
\author{Kartik P. Iyer\inst{1} \and Irene Mazzitelli\inst{1} \and
Fabio Bonaccorso\inst{1} \and 
Annick Pouquet\inst{2,3} \and Luca Biferale\inst{1}
}
%
%
\institute{University of Rome and INFN, Tor Vergata Rome, 00133, Italy
\and
Laboratory for Atmospheric and Space Physics, University of Colorado at 
Boulder, CO 80303, USA
\and 
Institute for Mathematics Applied to Geosciences (IMAGe), 
CISL, NCAR, Boulder, CO 80307-3000, USA}
\date{\textbf{Postprint version of the manuscript published in Eur. Phys. J. E, 
$38$ $12$ $(2015)$ $128$}}
%
\abstract{
The effects of changing the orientation of the rotation axis on 
homogeneous turbulence is considered. We perform
direct numerical simulations on a periodic box of $1024^3$ 
grid points, where the orientation of the rotation axis is changed
(a) at a fixed time instant (b) regularly at time intervals
commensurate with the rotation time scale. The former is characterized
by a dominant inverse energy cascade whereas in the latter,
the inverse cascade is
stymied due to the recurrent changes in the rotation axis resulting in a
strong forward energy transfer and large scale 
structures that resemble those of isotropic turbulence.
\PACS{
      {PACS-key}{discribing text of that key}   \and
      {PACS-key}{discribing text of that key}
     } 
} 
\maketitle
\section{Introduction}
\label{intro}
Turbulence subjected to rotation is a commonly occurring phenomenon
in geophysical and astrophysical flows, such as those in the
oceans and atmospheres of planetary bodies. 
The effects of rotation on decaying homogeneous
turbulence is typically studied by considering either an  isotropic or anisotropic state as the initial condition \cite{camb97,mans91}. 
In direct numerical simulations, 
different external forcing mechanisms have also been considered in order to
vary the degree of anisotropy and the typical correlation time 
\cite{SM14,SM15} of the flow.   
In most cases, it is customary to fix the orientation of the rotation axis. 
The general consensus is that when rotation is strong enough, 
the forward energy cascade from the large scales to the
small scales is inhibited  and  an inverse cascade develops resulting in
a quasi-two-dimensional behavior characterized by columnar structures
along the fixed rotation axis \cite{Pouquet2010,sen12}. 
The dominance of the inverse energy cascade entails the presence of a 
large scale energy sink in order to reach stationarity \cite{xia11}.
In this work, we
consider the effects of instantaneous changes to the orientation
of the rotation axis on homogeneous turbulence.

The interest in the orientation of rotation axis
is engendered by the phenomenon of precession, 
which is the rotational motion of the spin axis of 
a rotating body \cite{kida11}.
{
Even a weakly precessing container is known to
sustain turbulence at a sufficiently large Reynolds number,
due to viscous forces on the container 
walls \cite{goto07,Malkus}.
} 
For instance, the turbulent convection of
liquid metals in the Earth's outer core is influenced by its slow
precession. 
However details of the flow structure and 
the energy transfer dynamics in turbulence subjected to precession 
remain unclear
partly due to the fact that experiments as well as simulations
are difficult to conduct \cite{goto14}. 

As a recourse, we compare the spectral transfer properties of a system
perturbed by changing the orientation
of the axis at a given time instant 
and consequently allowed to relax,
with another system 
which is perturbed at regular time intervals by repeatedly 
changing the orientation
of the axis. We show that the latter has different
large scale properties and transfer dynamics 
as compared with the former. The remainder
of this work is organized as follows. In Sec.~\ref{sec2} we briefly
review the equations involved, simplifying assumptions made
and details about the simulations
performed. Results are given in Sec.~\ref{sec3}, with three 
subsections which focus respectively on (\ref{sec3a}) the evolution
of the kinetic energy and dissipation rate, (\ref{sec3b})
energy spectra and flux and (\ref{sec3c}) large scale structure evolution.
In Sec.~\ref{sec4} we summarize our results and discuss briefly the 
possible implications of this work.

\section{Numerical method}
\label{sec2}
The fluctuating velocity
$\uu(\ux,t)$ for a constant density flow in the co-ordinate system 
rotating with angular 
velocity $\omgv \equiv \omgv (t)$ is given by \cite{gspan}
\beqa
\nonumber
\mkern-18mu \!\!\!
\frac{\dho \uu}{\dho t} + \uu \cdot \unab \uu  +  2\omgv\times \uu +
\frac{d (\omgv \times \ur)}{dt} & = & \\  
\label{ns.eq}
 -\unab p + \mbf{f}  + \nu \delv \uu  - \gamma\delv^{-1}\uu \;, 
 & &
\eeqa
where $\mbf{f}$ is the 
forcing stirring the fluid, 
$\nu$ is the constant viscosity, $\gamma$ is the large scale
damping constant needed to remove energy at large scale and $\delv$ denotes the Laplacian operator.
The distance of the fluid particle from the 
rotation axis which precesses around a fixed axis is $|\ur|=r$.
In Eq.~\ref{ns.eq} the pressure $p$ accounts for the
centrifugal acceleration $\omgv \times
(\omgv \times \ur)$ in the usual manner.
In Eq.~\ref{ns.eq}, the precession term 
$d(\omgv \times \ur)/dt \sim \Omega L/\tau_p$, where $\tau_p$
is the precession time scale.
If $\tau_p$ is
large enough,
the precession term is negligible compared to the (time dependent) 
Coriolis term
since
$\Omega L/\tau_p \ll \Omega L/T_E$
where $L$ and $T_E$ denote the integral scale and the 
large eddy timescale of the flow.
Neglecting the precession term assuming
$\tau_p \gg T_E$ is a reasonable approximation in many
geodynamo applications which are characterized by large precession
time periods
\cite{nore,triana}. Another relevant scenario is that of a sudden change
in the orientation of the 
rotation axis at a given time instant, say at $t=0$, 
as a  sort of an instantaneous perturbation. 
In this case, 
one might expect that neglecting the non-homogeneous term becomes less and less important for the late-time dynamics, i.e. for $t \gg 0$. 
Accordingly, we solve the following equation numerically under the assumption that
the precession term $d(\omgv \times \ur)/dt$ can be neglected:
\beq
\label{nsolve.eq}
\frac{\dho \uu}{\dho t} + \uu \cdot \unab \uu + 2\omgv \times \uu
= 
-\unab p +\mbf{f} + \nu \delv \uu - \gamma\delv^{-1}\uu  \;.
\eeq
{
Equation \ref{nsolve.eq} is valid for weakly precessing flows which
are characterized by a large precession time scale $\tau_p$.
Alternatively, for sub-volumes of the flow close to the rotation axis,
such that
$|d(\omgv \times \ur)/dt| \sim \Omega r/\tau_p \to 0$
as $r \to 0$, 
the precession
term can be neglected in comparison to the Coriolis term.} 
The aim of this paper is to understand the evolution of the 
rotating flow under such sudden changes to the orientation of the rotation axis.
It also allows us 
to assess the robustness and universality of the large scale structures 
in strongly rotating turbulence.
We solve Eq.~\ref{nsolve.eq} using a 
Fourier pseudo-spectral method for spatial
discretization and a second order
Adams-Bashforth scheme for the time integration.
The domain is a
periodic cube with edge length
$L_0 = 2\pi$ with $N$ grid points to a side. The smallest
wave number in the domain is $k_0 = 1$.
The stochastic forcing applied to a shell
around the forcing wave number $k_f/k_0 = 4$ is based on a 
second-order Ornstein-Ulhenbeck process \cite{saw91}.
The hypo-viscous mechanism used to damp the large scales
is applied to wave numbers $k \in [0.5,2.5]$, with a 
large scale damping constant $\gamma = 0.1$ (refer Eqs.~\ref{ns.eq} and
\ref{nsolve.eq}).
Simulations wherein energy was depleted from different large scale ranges
were
also performed to test if the large scale properties
systematically depend
on the particular choice of wave numbers at which the energy was removed.
This issue of choice of low wave numbers from which energy is depleted
is further discussed in \ref{sec3c}.   
Aliasing errors from the nonlinear
term are effectively controlled by removing all coefficients
with wave number magnitude greater than $\kmax/k_0 = N/3$.
\section{Results and Analysis}
\label{sec3}
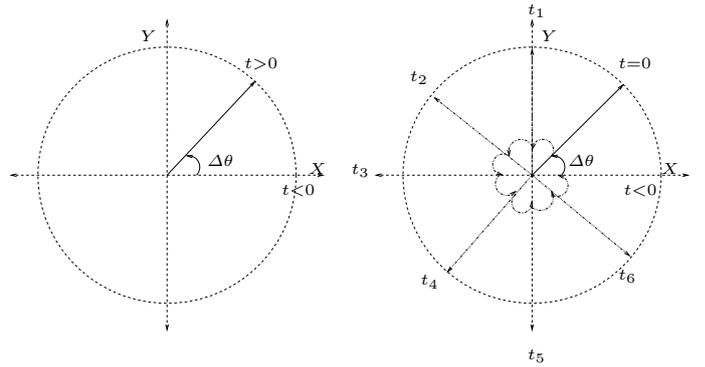
\begin{figure}
\center
\resizebox{0.5\textwidth}{!}{
\includegraphics{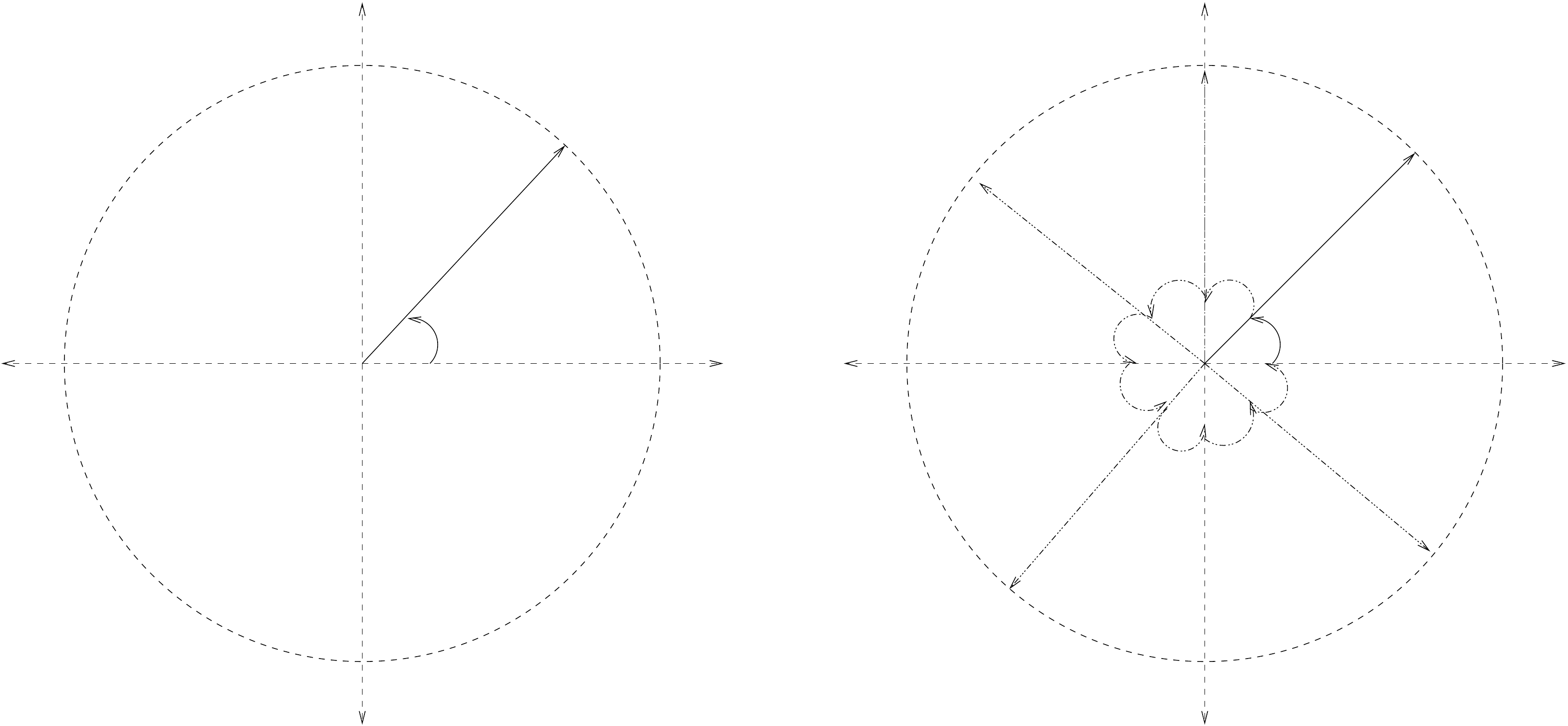}
}
\begin{picture}(1,1)
\put(106,62){$\ss{t < 0}$}
\put(-22,62){$\ss{t< 0}$}
\put(-50,73){$\ss{\Delta\th}$}
\put(85,73){$\ss{\Delta\th}$}
\put(-12,70){$\ss{X}$}
\put(120,70){$\ss{X}$}
\put(104,110){$\ss{t=0}$}
\put(-36,110){$\ss{t > 0}$}
\put(70,130){$\ss{t_1}$}
\put(26,105){$\ss{t_2}$}
\put(4,70){$\ss{t_3}$}
\put(30,28){$\ss{t_4}$}
\put(70,0){$\ss{t_5}$}
\put(104,30){$\ss{t_6}$}
\put(75,120){$\ss{Y}$}
\put(-75,120){$\ss{Y}$}
\end{picture}
\caption{Schematic of the orientation of the rotation vector
$\omgv$ in the $X$-$Y$ plane for simulations 
(left) R1 and (right) R2, $\Delta \th = \pi/4$. In run R1
$\omgv$ is rotated only at $t=0$ while in run R2 it is rotated 
at $t_i/\tomg = 10 i$, $i=0,1,2,\ldots$
}
\label{omgv.fig}
\end{figure}
We present results from two direct numerical simulations,
at grid resolution $1024^3$ and constant rotation magnitude $\Omega = 10$.
The rotation vector $\omgv$ lies in the $X$-$Y$ plane and is defined
by the angle $\th$ it makes with the $X$-axis (Fig.~\ref{omgv.fig}). 
A rotating, stationary
flow at $1024^3$, $\omg=10$, $\th=0$ 
is used as the initial condition ($t = 0$)
for both the simulations.
Statistical stationarity for $t \le 0$ is achieved using the
same large scale friction mechanism that is used in
simulations R1 and R2. 
\begin{table}
\centering
\caption{Initial and final values of Rossby numbers and related parameters
in the $1024^3$, $\Omega=10$ simulations.
The forcing wave number $k_f/k_0 = 4$. Large scale
dissipation is applied to wave number 
$k \in [0.5,2.5]$. 
Notes:
(1) $\tinit \equiv K(0)/\epsilon(0)$ is the turbulence
time scale at $t=0$. 
(2) $\tau_{\eta} = (\nu/\epsilon)^{1/2}$ is the
Kolmogorov time scale and $\eta = (\nu^3/\epsilon)^{1/4}$ is
the Kolmogorov length scale. 
}
\label{dns.tab}
\begin{tabular}{c|c|c|c}
\hline\noalign{\smallskip}
 & $t=0$ & R1 & R2 \\
\noalign{\smallskip}\hline\noalign{\smallskip}\\
$\tnr$ & $0.0$ & $11.75$ & $6.95$\\ 
$ \ro = \epsilon/(2K\Omega)$ & $0.0063$ & $0.0050$ & $0.0218$\\ 
$ Ro^{\omega} = 1/(2\tau_{\eta}\Omega)$ & $1.4231$ & $1.3151$ & $1.6863$\\
$\komgnr$ & $0.2083$ & $0.2344$ & $0.1615$\\  
\noalign{\smallskip}\hline
\end{tabular}
\end{table}
Simulation
R1 is performed using $\th=\pi/4$, while R2 is performed by 
instantaneously incrementing
$\th$ by $\Delta \th = \pi/4$ every $10 \tomg$, where
$\tomg \equiv 1/\Omega$ is the
rotation time scale. A schematic of the orientation of
$\omgv$ in the two runs is depicted in Fig.~\ref{omgv.fig}.
A conventional
measure of the strength of the rotation is given by comparing the rotation
time scale
to the turbulence time scale ($K/\epsilon$),
giving the turbulent Rossby number $Ro_T \equiv \epsilon/(2K\Omega)$. Here
$K$ is the kinetic energy and 
$\epsilon$ is the
mean dissipation and are defined in
Eqs. \ref{tke.eq} and \ref{diss.eq} respectively.
The definitions of the Rossby numbers including
initial and final values of these quantities
are summarized in Tab.~\ref{dns.tab}.
The initial large eddy time scale is defined as $\tinit \equiv K/\epsilon$ at
$t=0$. 
The Rossby numbers for run R2 at the end of the simulation
show an increase compared to their initial values
showing that the effect of rotation has possibly decreased.
{In contrast, for run R1 the turbulent Rossby number $Ro_T$
and the micro-Rossby number $Ro^{\omega}$ have decreased
indicating that effects of rotation are still significant.}
A characteristic wave number of rotation ($\komg$)
which delimits the region of the spectrum where rotation
effects are important ($k < \komg$) is given by
$\komg = ({\omg}^3/\epsilon)^{1/2}$ \cite{zee94}.
The non-dimensional rotation wave number
$\komgnr \sim 1$ in the simulations (Tab.~\ref{dns.tab}),
indicating that the rotation effects extend to the viscous
scales in the flow.
\subsection{Energy and dissipation}
\label{sec3a}
\begin{figure*}
\resizebox{1.0\textwidth}{!}{
\includegraphics{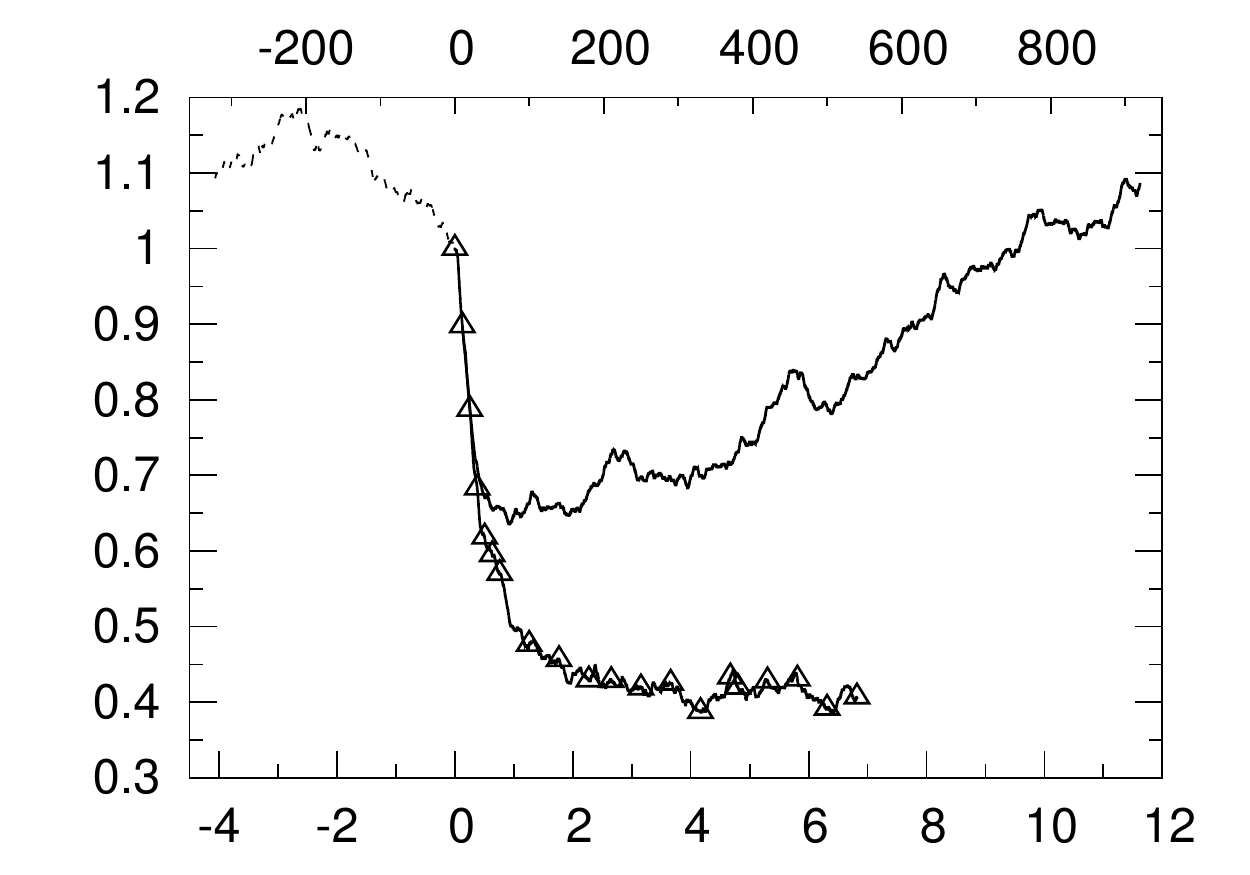}
\includegraphics{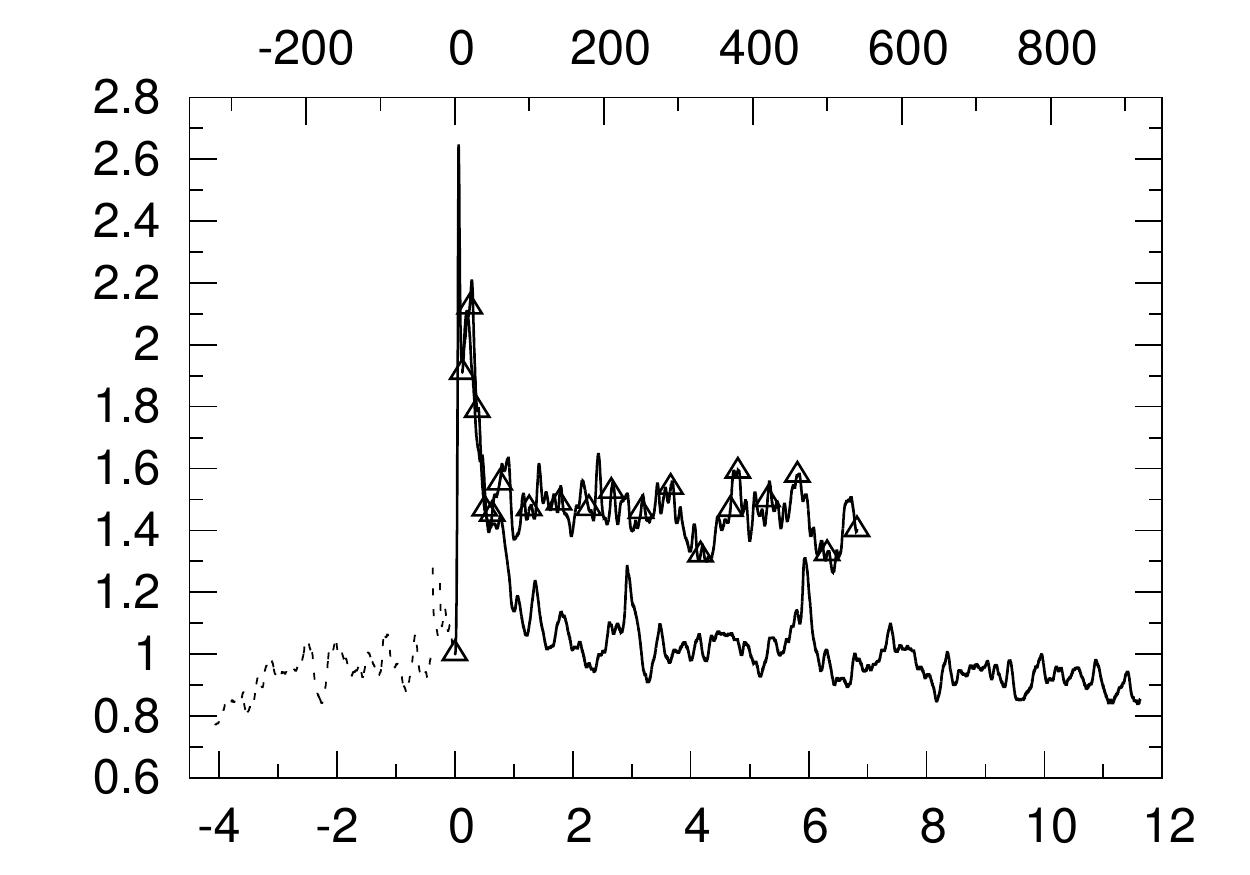}
}
\begin{picture}(1,1)
\put(120,2){$\tnr$}
\put(120,187){$\tomgnr$}
\put(380,2){$\tnr$}
\put(380,187){$\tomgnr$}
\put(3,70){\rotatebox{90}{$K(t)/K(0)$}}
\put(255,70){\rotatebox{90}{$\epsilon(t)/\epsilon(0)$}}
\end{picture}
\caption{Evolution of kinetic energy (left) and
dissipation (right) in normalized time,
compared with respective values at $t=0$.
Solid curves corresponds to run R1, curves with symbols ($\bigtriangleup$)
correspond to run R2. Dashed curves show 
histories of $K(t)$ and $\epsilon(t)$ prior
to start ($t < 0$) of the perturbation. 
Time axis is normalized by
large eddy time scale at $t=0$ ($\tinit$) in bottom horizontal 
axis and by rotation time scale ($\tomg$) in top horizontal axis.
}
\label{eediss.fig}
\end{figure*}
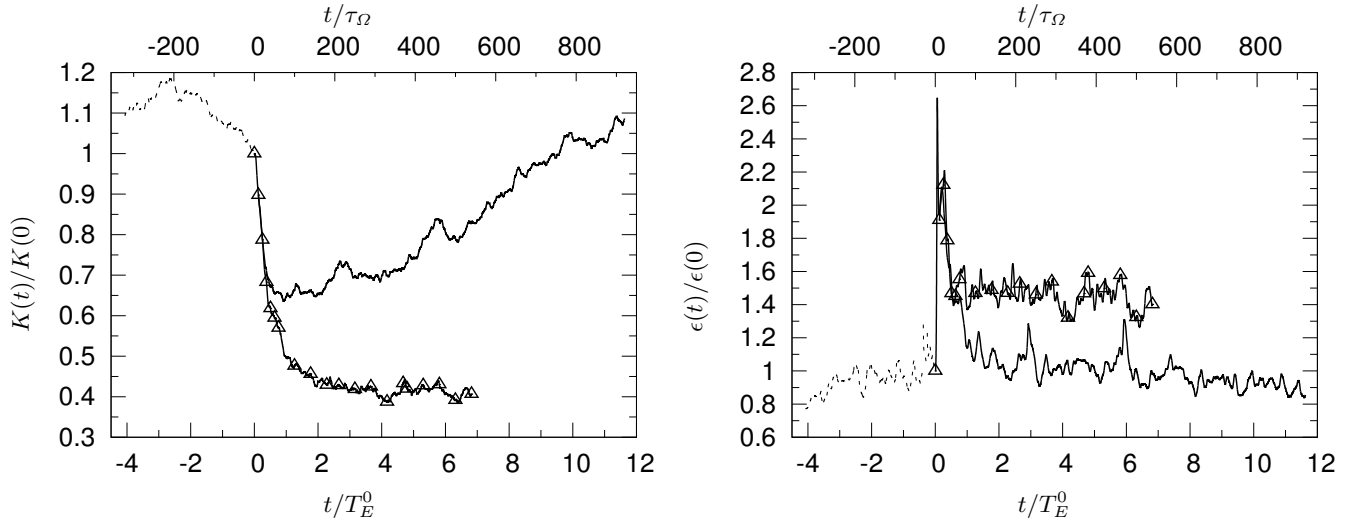
The mean turbulent kinetic energy and dissipation provide important global
measures of the state of the turbulence under rotation. In homogeneous
turbulence, the mean turbulent kinetic energy is 
defined as
\beq
\label{tke.eq}
K = \frac{1}{2} \bla \sum_{i=1}^3 u_i^2 \bra \;,
\eeq
and the mean dissipation rate is given by
\beq
\label{diss.eq}
\epsilon=2\nu \bla \sum_{i,j=1}^3 \frac{\dho u_i}{\dho x_j} \frac{\dho u_j}{\dho x_i}\bra\;,
\eeq
where,  
$\langle \cdot \rangle$ denote space averages. 
In Fig.~\ref{eediss.fig}
we show the evolution of these quantities normalized by their
initial values.
Despite the
considerable statistical variability,
it is clear that the kinetic
energy initially decreases for simulation R1, 
but subsequently
increases with time. On the other hand, the energy
for simulation R2 drops in the early stages and then remains nearly
constant at a value well below that of R1. 
The steep drop in the energy
at early times is accompanied by a sharp increase in
mean dissipation. The dissipation rates then drop
to nearly constant values with that of simulation R2 stabilizing
at a value larger than that of R1.
Furthermore, the time histories of energy and dissipation prior
to the start of the perturbation 
($t < 0$ in Fig.~\ref{eediss.fig})
indicate that the initial changes in energy and dissipation caused by the
change in orientation of $\omgv$ are
significant. The previous observations indicate that a single perturbation in the
form of an 
instantaneous change to the orientation of the rotation axis
results in a slow recovery of the 
``universal'' 
inverse energy transfer mechanism. In contrast, if the system is 
subjected to such perturbations repeatedly, the
inverse energy transfer is eventually destroyed as the dynamical 
reconstruction of the large scale structures is too slow to survive. 
Indeed, the dissipation of run R2 attains 
a quasi-stationary state which is higher than its initial value,
while its energy stabilizes at a lower value.
This indicates a net energy transfer 
from large scales to the small scales in simulation R2.

\begin{figure*}
\resizebox{1.0\textwidth}{!}{
\includegraphics{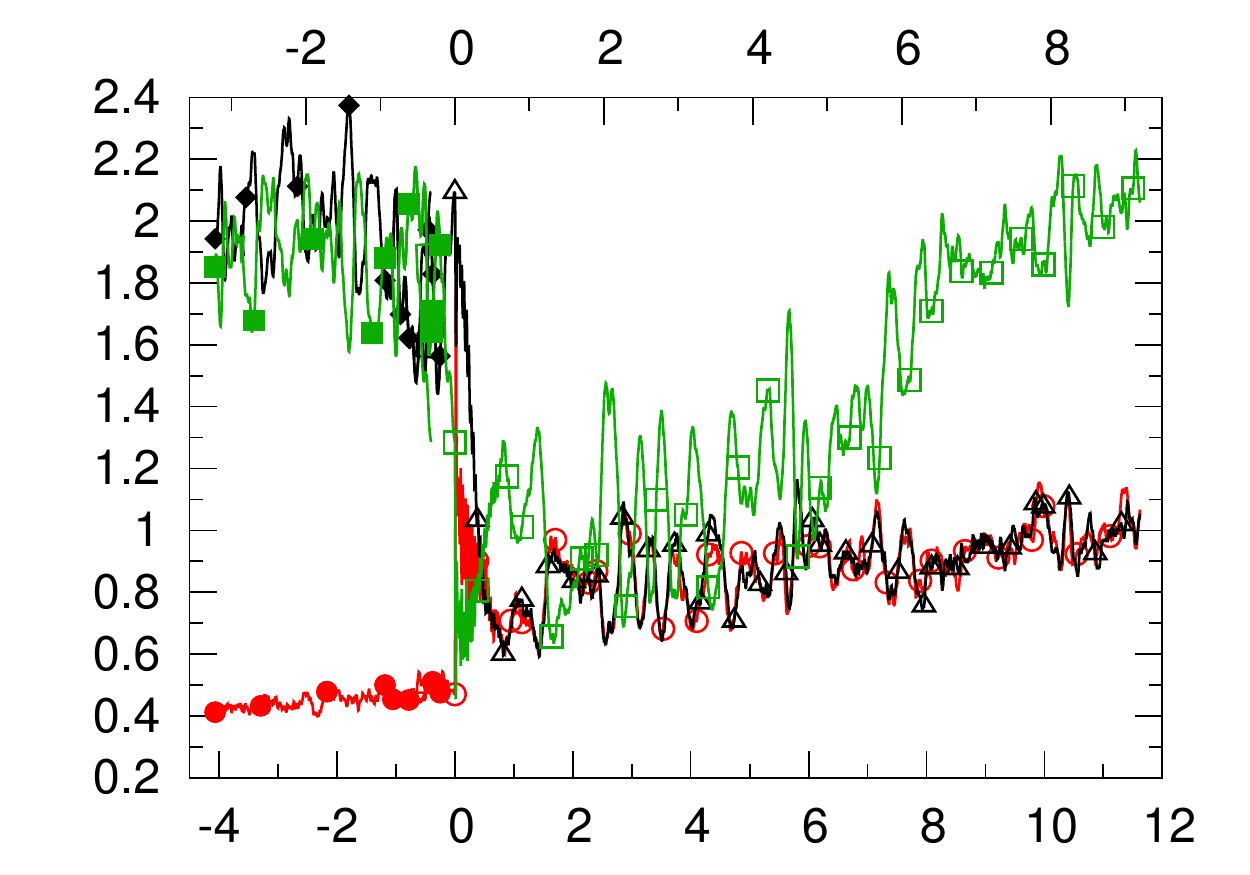}
\includegraphics{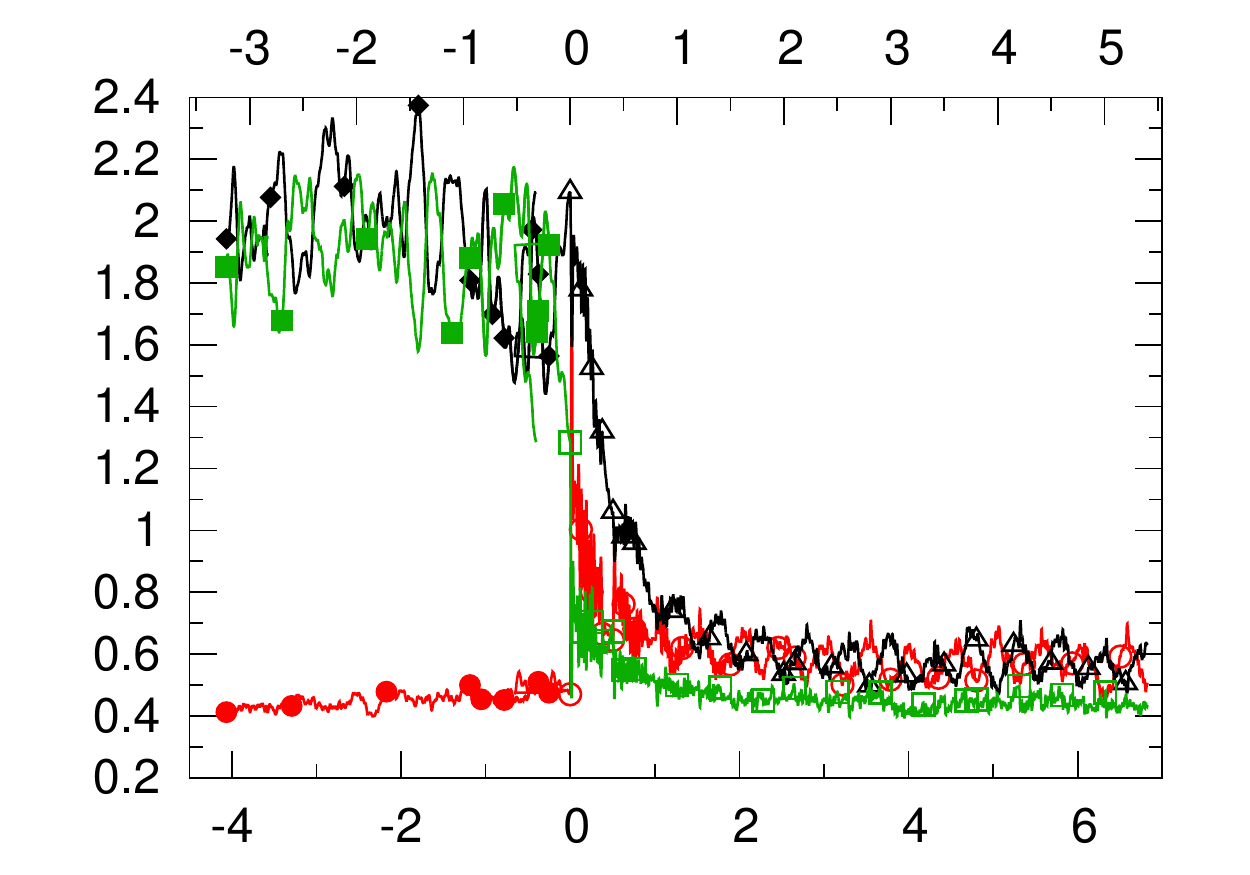}
}
\begin{picture}(1,1)
\put(120,2){$\tnr$}
\put(120,187){$(\tomgnr){10}^{-2}$}
\put(380,2){$\tnr$}
\put(380,187){$(\tomgnr){10}^{-2}$}
\put(3,70){\rotatebox{90}{$\la u_\alpha^2 \ra/2$}}
\end{picture}
\protect\caption{(Color online) Evolution of the three components of the turbulent kinetic energy
$\la u_\alpha^2 \ra, \; \alpha=1,2,3$ as a function of normalized time
for run 
(left) R1 and (right) R2. Open symbols ($\Circle$), ($\bigtriangleup$) and 
($\square$) 
correspond to $\la u_1^2 \ra$, $\la u_2^2 \ra$ and $\la u_3^2 \ra$
respectively in the simulations.
Closed symbols ($\CIRCLE$), ($\blacklozenge$) and ($\blacksquare$) correspond
to time histories ($t < 0$) of $\la u_1^2 \ra$, $\la u_2^2 \ra$ and $\la u_3^2 \ra$
respectively prior to the start of the perturbation.
Time axis is normalized by
large eddy time scale at $t=0$ ($\tinit$) in bottom horizontal 
axis and by rotation time scale ($\tomg$) in top horizontal axis.
}
\label{vvar.fig}
\end{figure*}
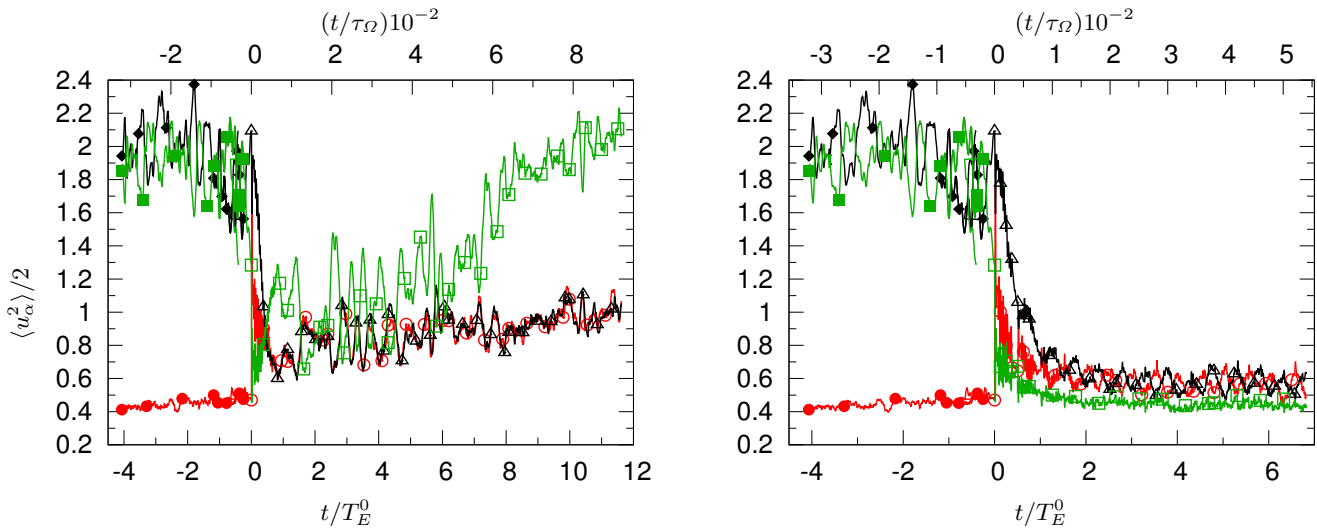
The evolution of the Cartesian components of the turbulent kinetic
energy are indicative of large scale anisotropy in the flow.
Figure \ref{vvar.fig} compares the variance of the 
velocity components $\la u_\alpha^2 \ra$
in the two simulations. Prior to the start of the perturbations, 
the velocity fluctuations perpendicular to the rotation axis 
($\th = 0$ for $t<0$ in Fig.~\ref{omgv.fig}) are dominant due to 
an inverse energy cascade in the $Y$-$Z$ plane. The
instantaneous rotation of $\omgv$ at $t=0$
disrupts the spectral transfer to the largest scales in the $Y$-$Z$
plane. With time, in run R1 an inverse cascade in the plane normal to the
new rotation axis is established resulting in increasing
energy in the $Z$-direction. The variance of velocity components in
the $X$ and $Y$ directions evolve similarly at a lower value than 
the $Z$ component.  
Whereas in run R2, the disruption of the inverse cascade at
$t=0$ is sustained by the regular change in the orientation of the
rotation axis. As a result, the energy components reach a stationary isotropic 
state.
\vspace{-1.0em}
\subsection{Energy spectra and flux }
\label{sec3b}
\begin{figure*}
\resizebox{1.\textwidth}{!}{
\includegraphics{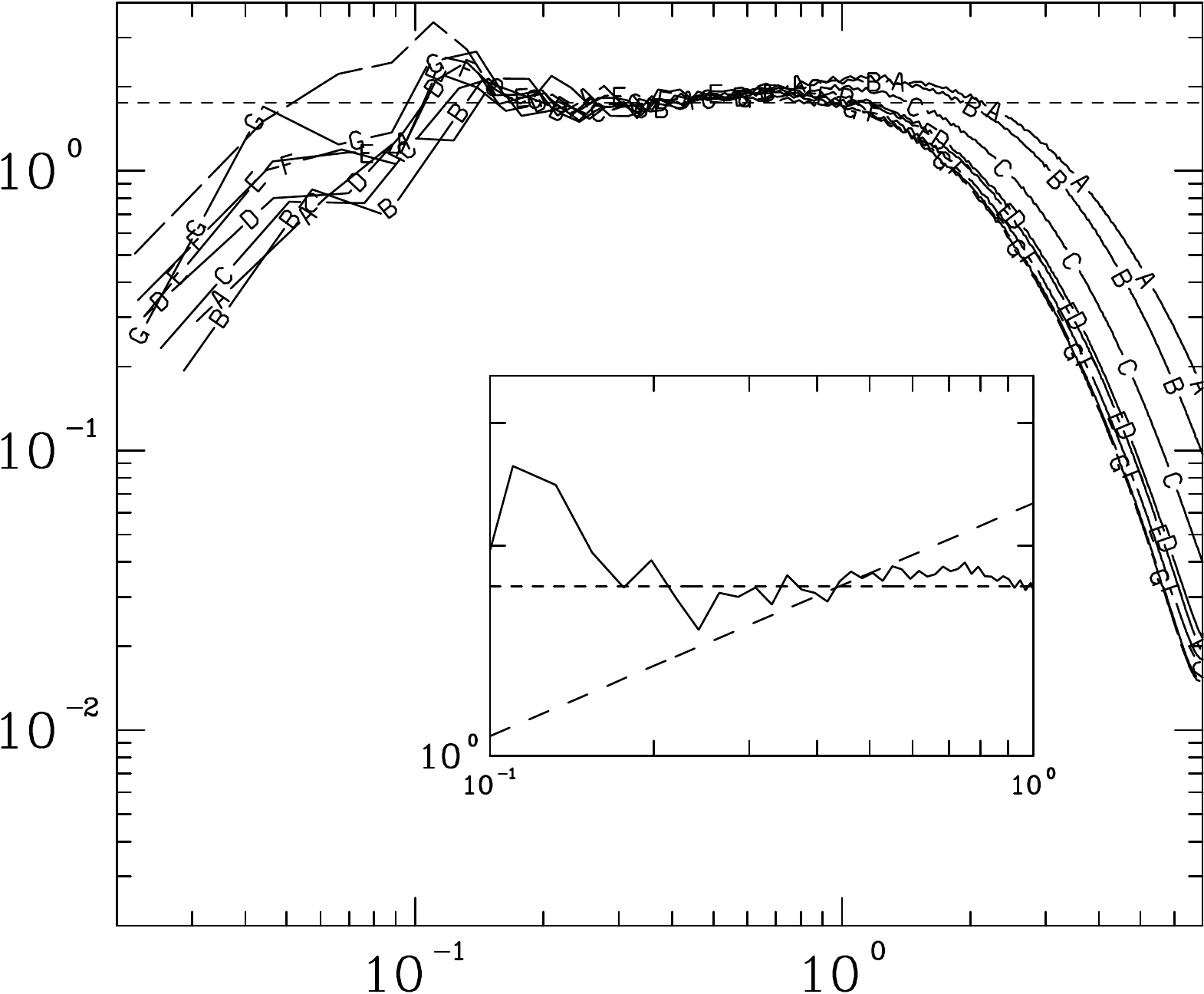}
\includegraphics{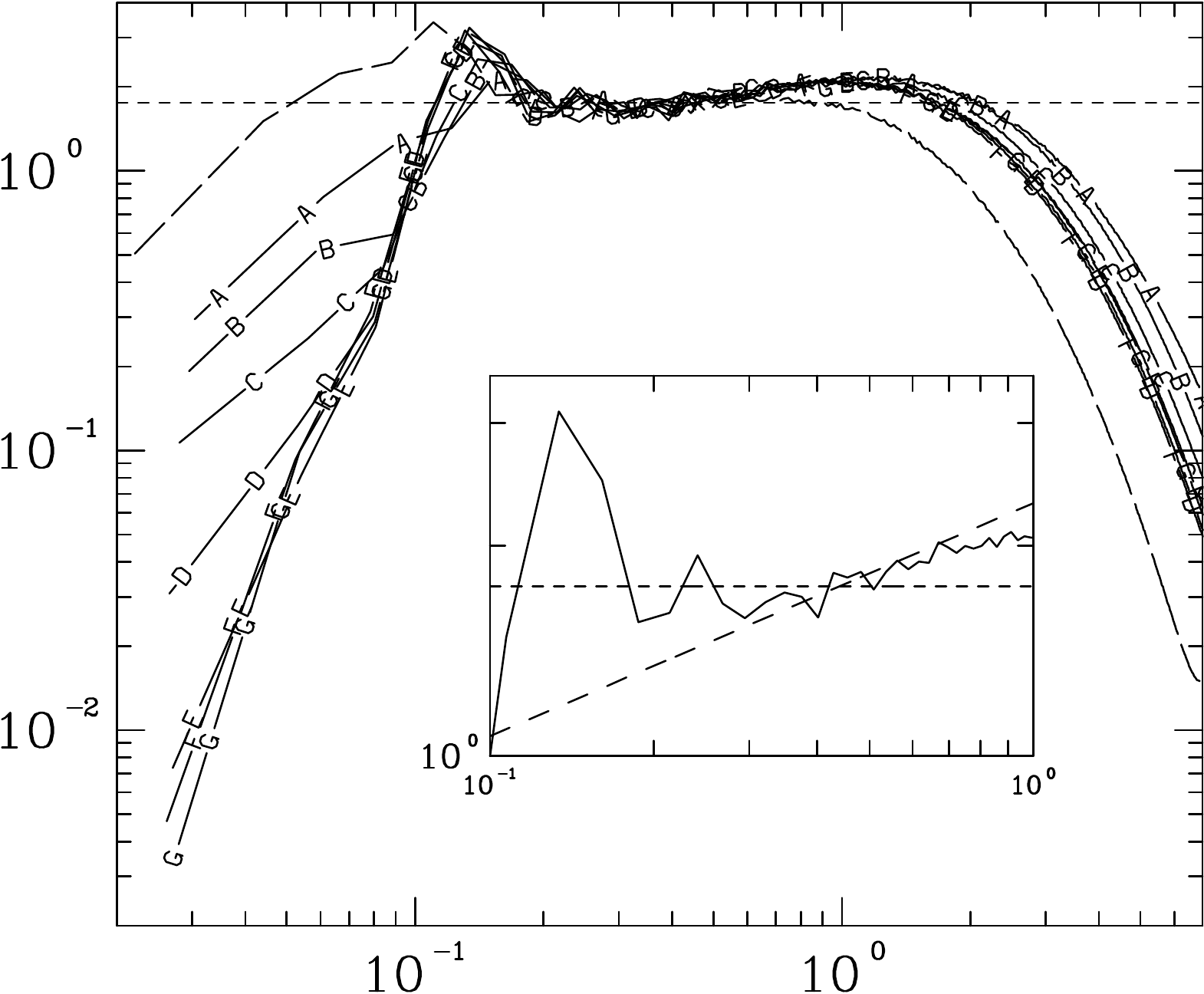}
}
\begin{picture}(1,1)
\put(120,2){$\knorm$}
\put(145,45){$\ss{\knorm}$}
\put(200,115){$\ss{k^{1/3}}$}
\put(68,47){$k_f/\komg$}
\put(380,2){$\knorm$}
\put(410,42){$\ss{\knorm}$}
\put(455,112){$\ss{k^{1/3}}$}
\put(325,47){$k_f/\komg$}
\put(89,25){\vector(0,1){20}}
\put(344,25){\vector(0,1){20}}
\put(-12,70){\rotatebox{90}{$E(k,t)(\epsilon(t)\Omega)^{-1/2}k^2$}}
\put(85,70){\rotatebox{90}{$\ss{E(k)(\epsilon\Omega)^{-1/2}k^2}$}}
\put(340,70){\rotatebox{90}{$\ss{E(k)(\epsilon\Omega)^{-1/2}k^2}$}}
\end{picture}
\protect\caption{Energy spectrum normalized by rotation scaling at
different times for (left) simulation R1 and (right) simulation R2.
Symbols $A$-$G$ correspond to $\tnr = 1/4,1/2,1,2,3,4,5$
respectively.
 Curve with (long) dashes corresponds to $t=0$.
Dashed horizontal lines at $C_\Omega = 1.75$ for reference.
Forcing wave number magnitude $k_f/\komg \approx 0.1 $ for
simulations R1 and R2 at all times shown. {Inset shows blow-up of the
intermediate scale range $k_f \le k \le \komg$ at late-time.  
Dashed horizontal line at $C_\Omega = 1.75$ corresponds to 
$k^{-2}$ scaling, while long-dashed
line corresponds to Kolmogorov $-5/3$ spectrum \cite{krs95}.} 
}
\label{ek.fig}
\end{figure*}
In the rotation-modified inertial range, theoretical arguments
\cite{zhou95} suggest
that for $k_f \ll k \ll \komg$  the
energy spectrum is of the form
\beq
\label{spec.eq}
E(k) = C_\Omega (\epsilon \omg)^{1/2} k^{-2}\;,
\eeq
where the constant $C_\Omega = 1.22-1.87$ \cite{zhou95}.
In Fig.~\ref{ek.fig} we show the development of the compensated
energy spectrum $E(k)(\epsilon \omg)^{-1/2} k^{2}$ at
different times for both simulations R1 and R2. The energy at low
wave numbers ($k < k_f$) initially decreases ($\tnr < 1$)
and then increases for run R1 
whereas the energy at the largest scales in run R2 monotonically
decreases with time. These results are
consistent with energy evolution trends of the two
simulations shown in Fig.~\ref{eediss.fig}. 
{At the intermediate 
wave numbers $k_f \ll k \ll \komg$,
run R1 exhibits a $k^{-2}$ behavior with an inertial range
plateau of $C_\omg = 1.75$.}
{
In contrast, run R2 shows a greater tendency towards 
a transition to the  classical $k^{-5/3}$ scaling in the inertial range
which is 
typical of flows without strong rotation \cite{min12}. 
On the other hand, the energy at the high wave numbers ($k > \komg$)
is greater in run R2 than in run R1,  indicating a 
stronger forward cascade in the former.}
The results
confirm that the
transfer dynamics in the case with ``precession-like'' perturbations
are inherently different than the case with a fixed rotation axis.

\begin{figure*}
\resizebox{1.0\textwidth}{!}{
\includegraphics{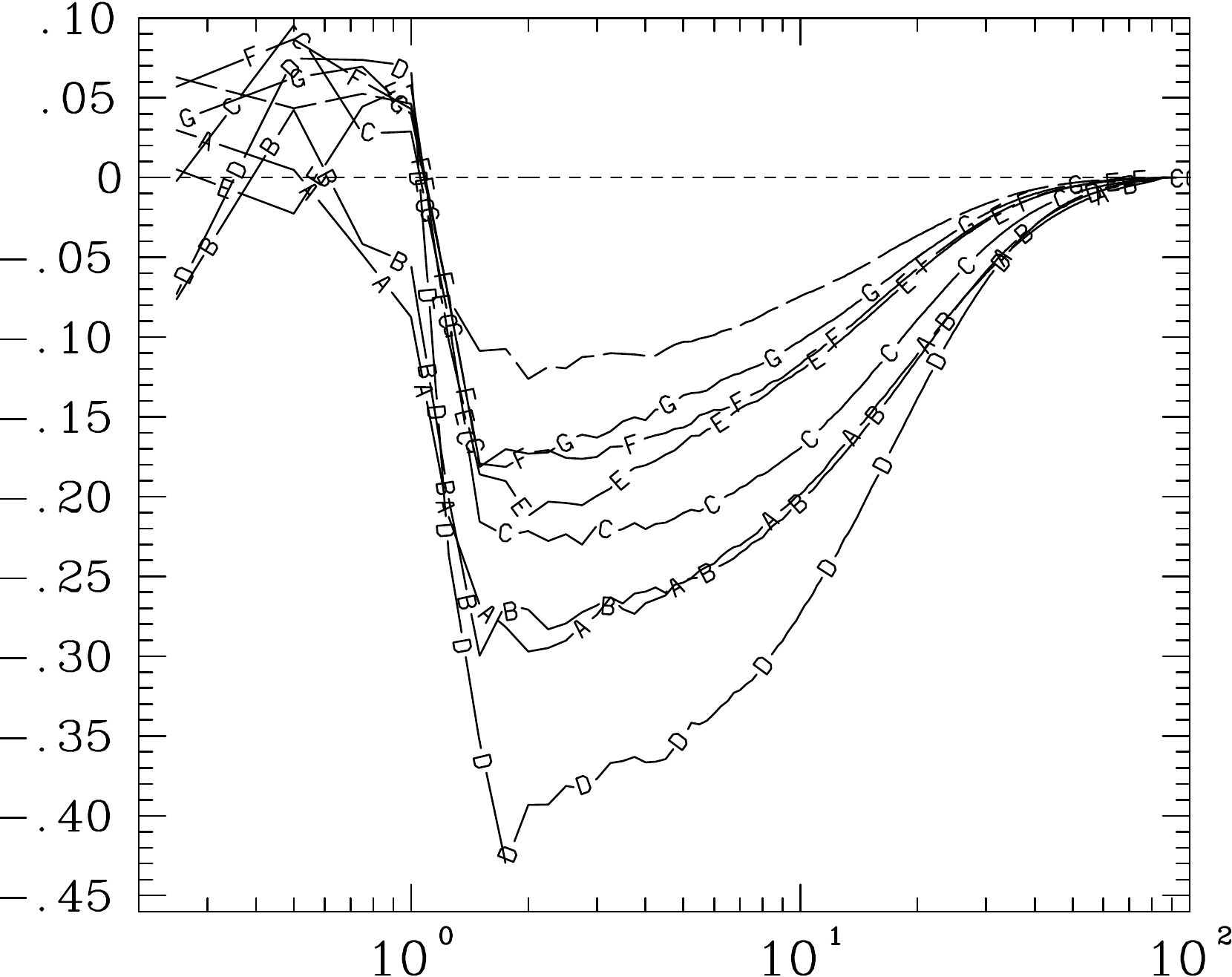}
\includegraphics{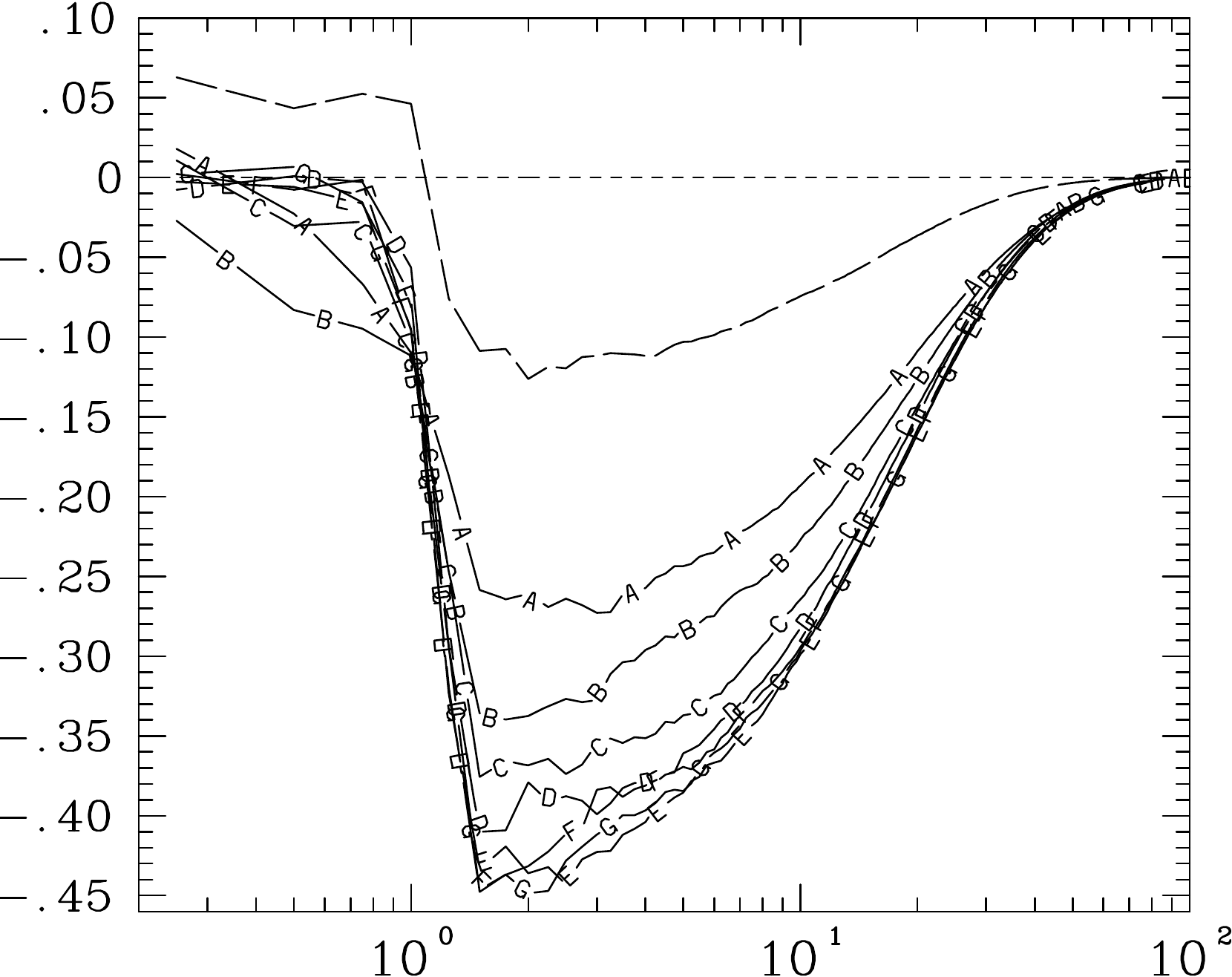}
}
\begin{picture}(1,1)
\put(120,165){$t=0$}
\put(124,163){\vector(1,-1){20}}
\put(120,2){$k/k_f$}
\put(380,165){$t=0$}
\put(384,163){\vector(1,-1){20}}
\put(380,2){$k/k_f$}
\put(-10,80){\rotatebox{90}{$\Pi(k,t)/K(t)$}}
\end{picture}
\caption{Spectral flux (Eq.~\ref{flx.eq}) normalized
by kinetic energy at
different times for (left) run R1 and (right) run R2.
Different curves correspond to different instants of time
(refer Fig.~\ref{ek.fig}).  
Dashed horizontal line at $0$ for reference.
The maximum (negative) amplitude of flux increases with time (curves $A$-$G$) for
run R2. Positive spectral flux corresponds to an inverse energy cascade while negative
flux indicates a forward cascade (see Eq.~\ref{flx.eq}). 
}
\label{flux.fig}
\end{figure*}
The direction of energy transfer can be conveniently studied by
examining the contribution of the nonlinear terms
in Eq.~\ref{nsolve.eq}
to the rate of change of energy in $k$-space. 
Following \cite{MY.II} we define the spectral
flux as
{ 
\beq
\label{flx.eq}
\Pi(k) = \!\! \int\dl_0^k \!{\textrm{Im}\bla{
\kd_m P_{ij}(\kdv)\uihat^{*}(\kdv)
\mkern-18mu \!\!\!
\int\dl_{\kdv=\mathbf{p}+\mathbf{q}}{
\mkern-18mu
\ujhat(\mathbf{p})
\hat{u}_m (\mathbf{q}) }}\bra d \mathbf{p} \; d\kd }.
\eeq
}
Here $\textrm{Im}(\cdot)$ denotes imaginary part of $(\cdot)$,
 overcarets represent Fourier coefficients, $(\cdot)^{*}$ is the
complex conjugate of $(\cdot)$ and the tensor
$P_{ij}(\kv)=k_i k_j/k^2 - \delta_{ij}$ represents projections
onto the plane perpendicular to 
$\kv$ in wave number space ($\delta_{ij}$ is 
the Kronecker delta tensor).
Figure \ref{flux.fig} shows the flux normalized by kinetic 
energy $K(t)$ at various times for both
simulations.
The positive plateau at low wave numbers ($k < k_f$)
accompanied by a decrease in the 
flux magnitude at higher
wave numbers ($k > k_f$)
at later times is a clear indication of large-scale
energy transfer in simulation R1.
Whereas, the low wave number flux is almost zero for simulation R2
at later times,
indicating that the inverse cascade is suppressed by the repeated
change in the rotation axis orientation.
Furthermore the flux magnitude
 at higher wave numbers ($k > k_f$) increases with
time in run R2 indicating that the 
perpetual change in the orientation of the rotation axis enhances the
forward energy cascade.

\begin{figure*}
\resizebox{1.0\textwidth}{!}{
\includegraphics{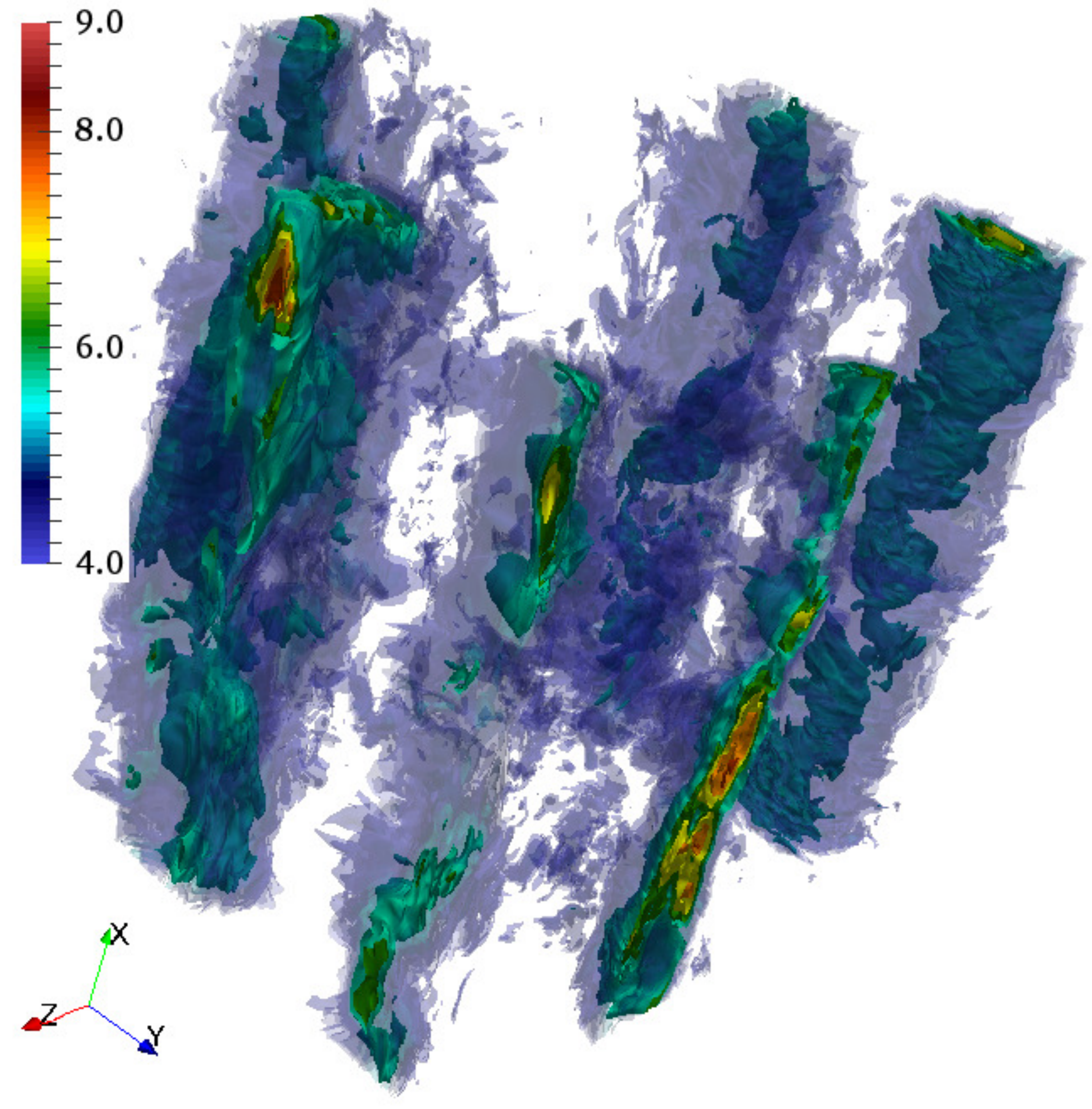}
\includegraphics{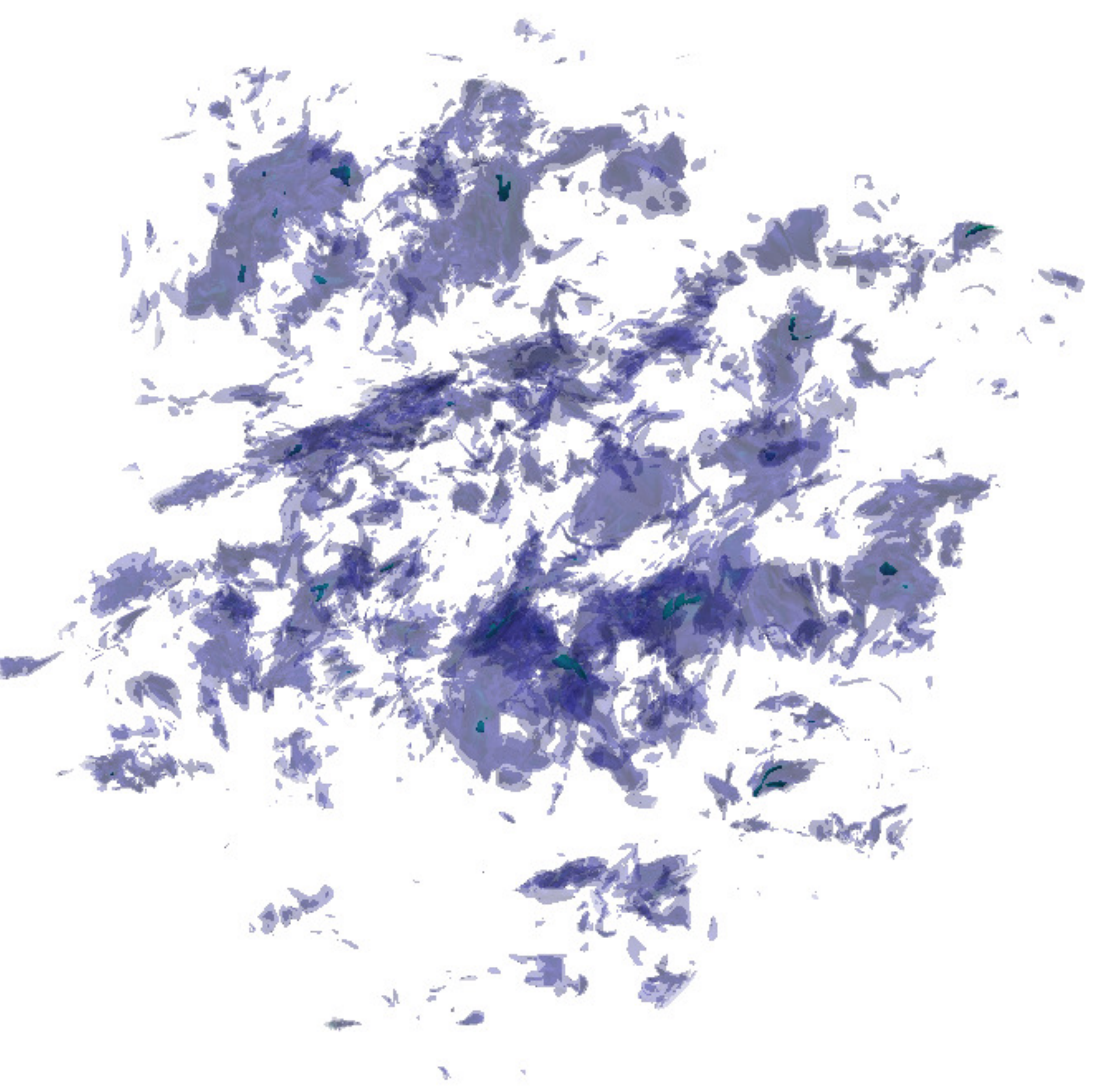}
\includegraphics{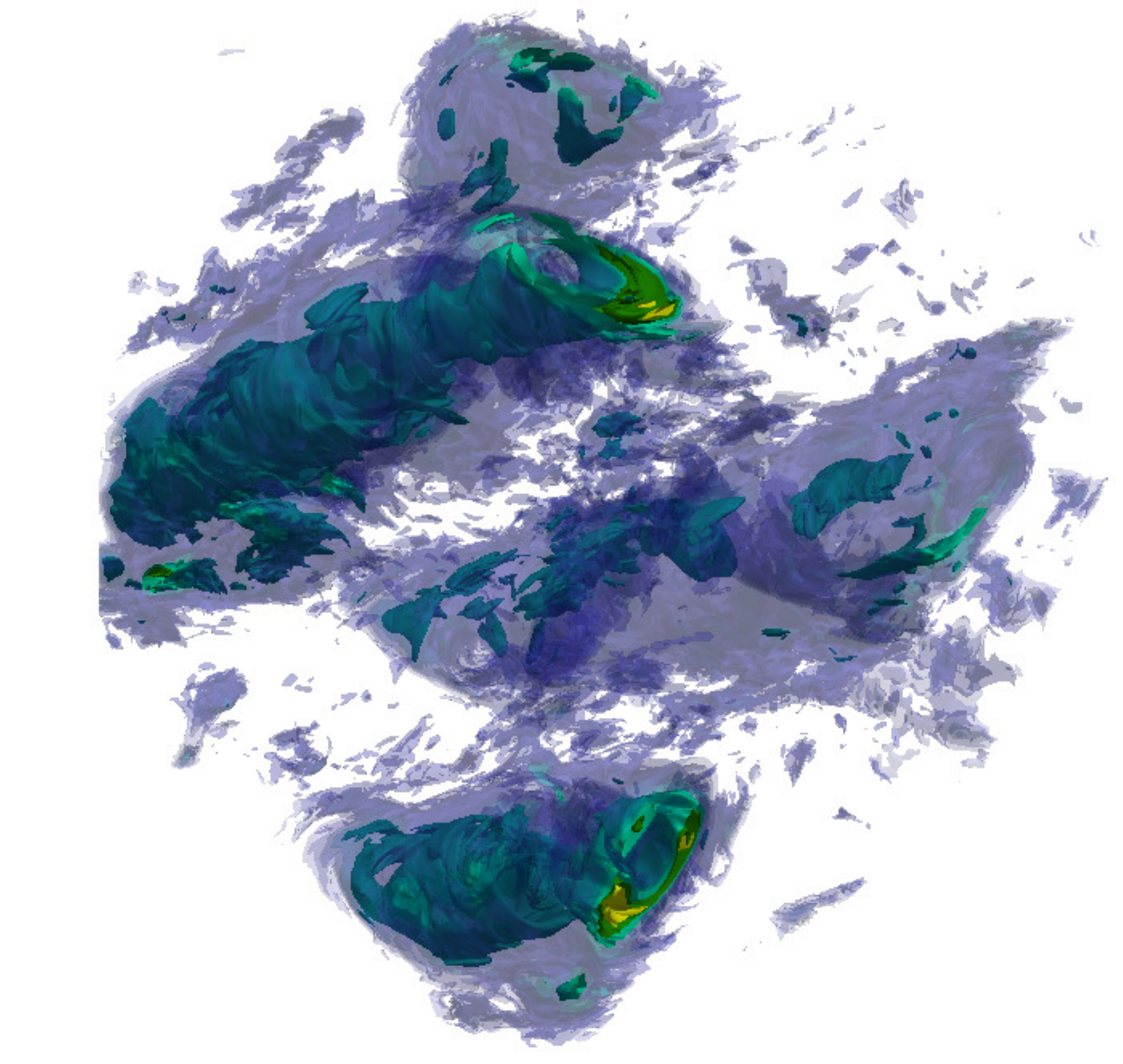}
}
\caption{
(Color online) Iso-contours of magnitude of velocity fluctuations
($\sqrt{K(t)}$) in run R1 for (left-right) $\tnr=0,1,11.5$.
}
\label{exp3.fig}
\end{figure*}
\begin{figure*}
\resizebox{1.0\textwidth}{!}{
\includegraphics{vizcropped_new/T1.pdf}
\includegraphics{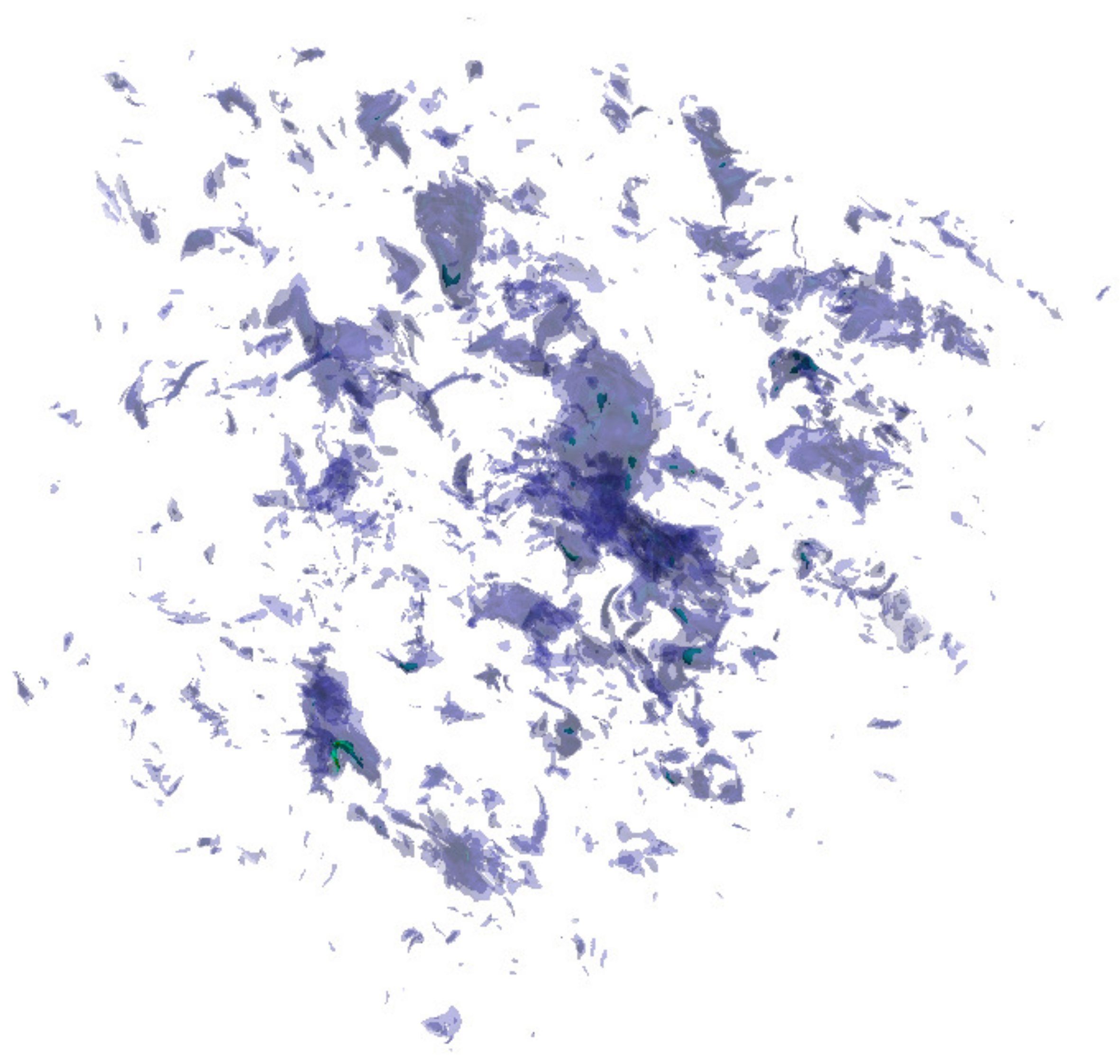}
\includegraphics{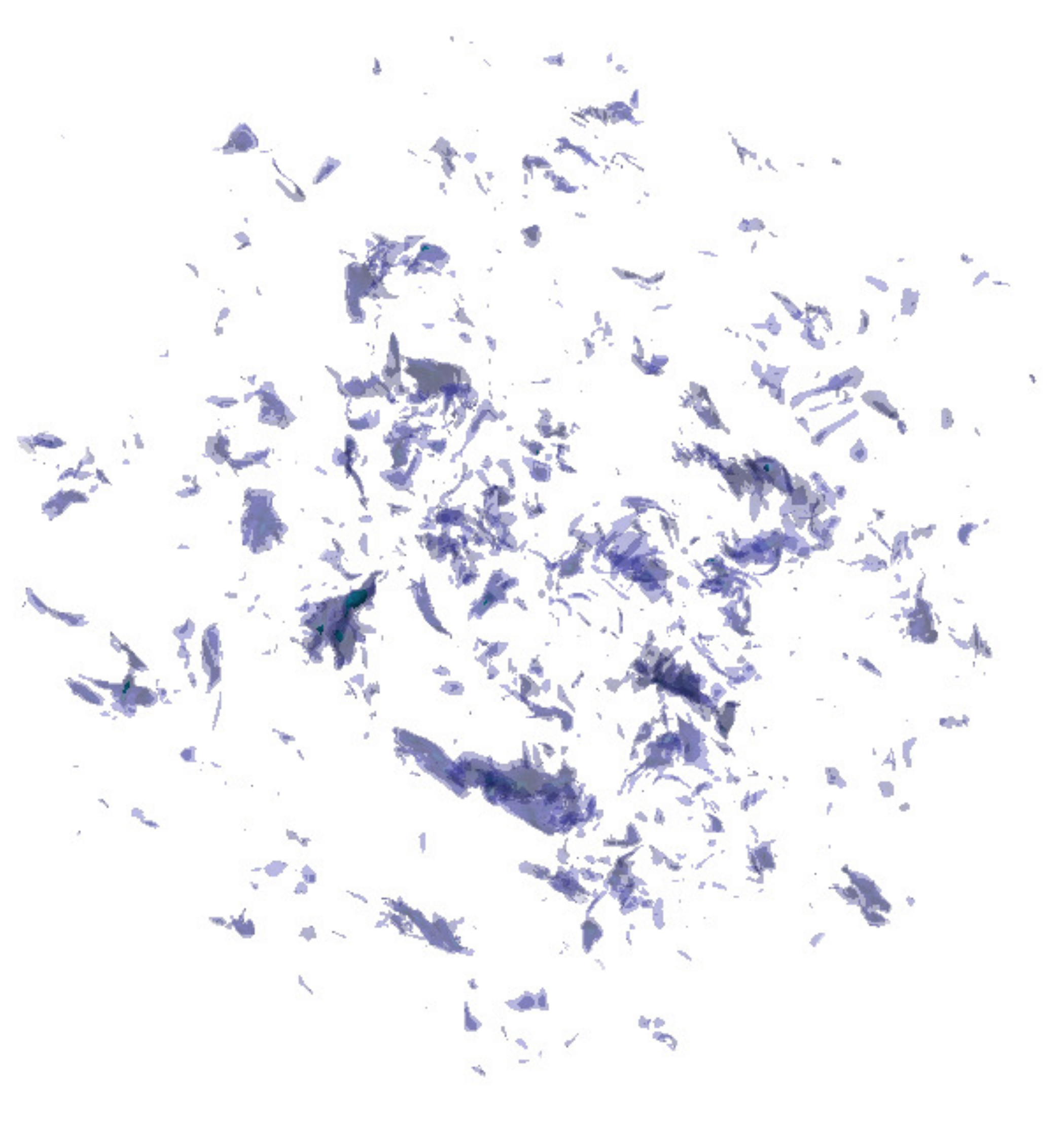}
}
\caption{
(Color online) Iso-contours of magnitude of velocity fluctuations
($\sqrt{K(t)}$) in run R2 for (left-right) $\tnr=0,1,6.95$.
Note that the iso-contours at $t=0$ for R2 are the same as that
for R1 at $t=0$ (left panel in Fig.~\ref{exp3.fig}) as both runs
start from same initial conditions.
}
\label{exp1.fig}
\end{figure*}
\subsection{Large scale structure}
\label{sec3c}
Independent of the realizability of the numerical experiments shown here, 
an important consideration in turbulence subjected to rotation is
the robustness and universality of the large scale structures under sudden perturbations 
of the large scale set-up \cite{SM14}. 
Figures \ref{exp3.fig} and \ref{exp1.fig} show the iso-contours of the
velocity magnitude at three different time instants in simulations
R1 and R2 respectively. For $t \le 0$, the inverse cascade in the 
plane normal to the rotation axis manifests itself as columnar structures
along the axis of rotation \cite{map09}.
Shortly after the change in the orientation of the rotation axis
at $t=0$, the inverse cascade is suppressed for both
runs R1 and R2 but their further evolution differ. 
In run R1 after a transient, the energy flux becomes positive again at the 
large scales and diminishes in magnitude at the small scales (Fig.~\ref{flux.fig}).
On the other hand, in run R2 the inverse cascade dynamics has insufficient time
to recover and the direct cascade is stronger. 
This can be illustrated by the time evolution of the maximum (negative)
amplitude of the flux which is non-monotonous only for run R1 (Fig.~\ref{flux.fig}).
The large scale structure evolution can be associated with the life-time 
of the individual eddies versus
the time required for the build-up of the forward cascade. At the 
low Rossby numbers considered,
the time between switching the rotation axis is comparable with the 
rotation time scale. In such a scenario, 
the inverse cascade does not have the time to rebuild as evidenced by the
lack of large scale structures for $t > 0$ in Fig.~\ref{exp1.fig}.  
Thus, the columnar structures characteristic of turbulence subjected
to rotation in a fixed direction
(Fig.~\ref{exp3.fig})  
are absent when the orientation of the rotation axis
is changed fast enough (Fig.~\ref{exp1.fig}). 
The velocity field structure in the case where the
rotation axes is regularly changed
resembles that of an isotropic field, a visual proof that the 
inverse energy cascade is not sustained.

\begin{figure}
\center
\resizebox{0.5\textwidth}{!}{
\includegraphics{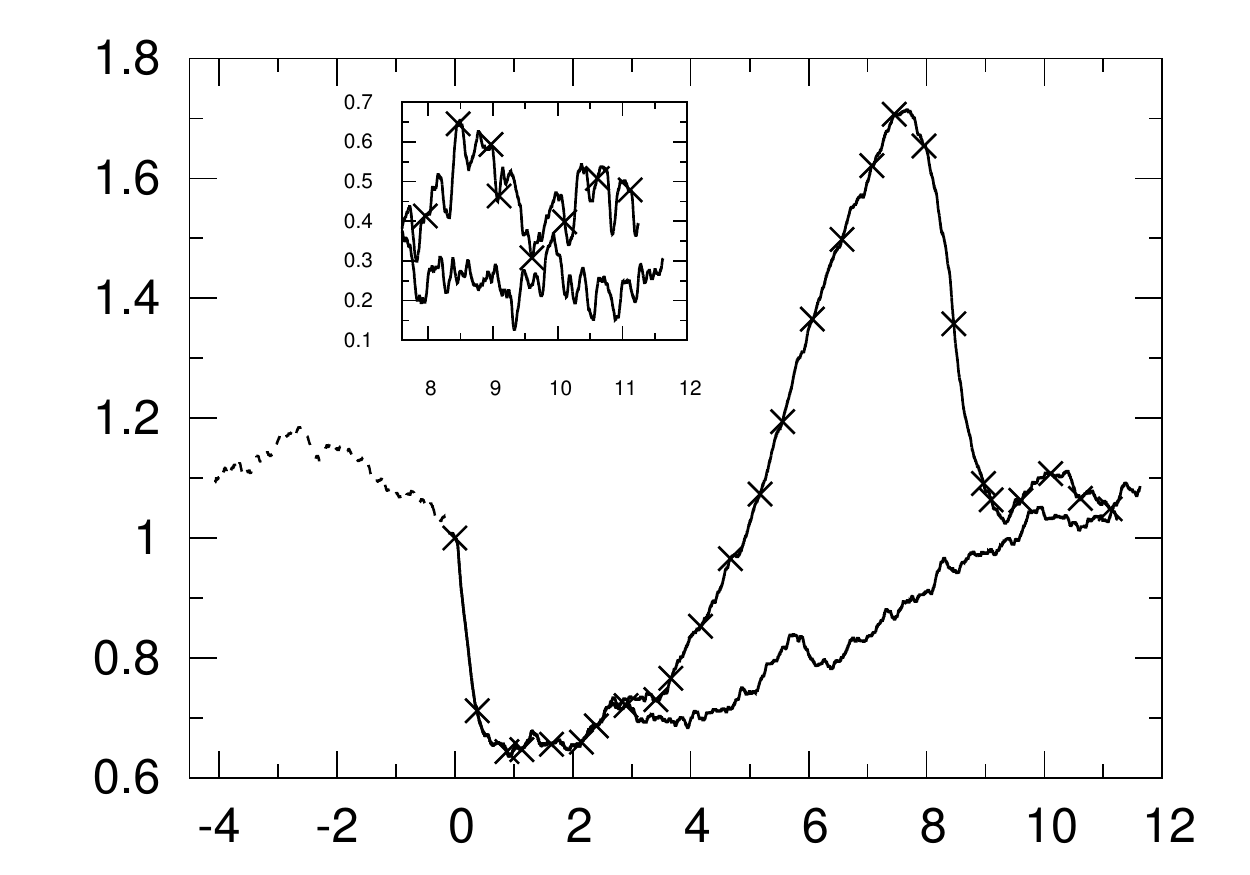}
}
\begin{picture}(1,1)
\put(-0,4){$\tnr$}
\put(-20,95){$\ss\tnr$}
\put(-120,80){\rotatebox{90}{$K(t)/K(0)$}}
\put(-70,120){\rotatebox{90}{$\ss{L_{21,1}/(\frac{1}{2}L_0)}$}}
\end{picture}
\caption{
Evolution of energy for simulation R1 for two different cases
of large-scale damping. Solid curve corresponds to energy removal
for wave number $k \in [0.5,2.5]$, while curve with symbols
($\times$) corresponds to $k \in [0.5,1.5]$. Inset shows
late-time evolution of transverse integral length scale 
$L_{21,1}$ normalized by half the length
of the domain ($L_0$).
}
\label{damp.fig}
\end{figure}

Another question we attempt to address is the robustness of 
the large scale structures as a function of the 
energy sink  mechanism applied at large scales. 
In simulations R1 and R2 a hypo-viscous mechanism
 is used to prevent 
energy condensation that can occur because of upscale energy transfer
owing to 
finite domain considerations in rotating flows \cite{xia11,SY93}. 
It is reasonable to
expect that the large scale statistics are influenced
by the details of the friction mechanism that is used at the low
wave numbers.
In order to verify this, 
we changed  the
energy removed from the largest scales. For instance, in 
run  R1 we removed energy from different
wave number shells at the large scales, 
thus depleting the system differently.
Figure \ref{damp.fig} shows two such scenarios where
the energy
is removed from shells $0.5 \le k \le 2.5$ and
$0.5 \le k \le 1.5$, respectively. Removing energy from a thinner shell
results in a steeper increase in energy initially, but ultimately the
energy decreases due to the action of large-scale viscosity 
\cite{cher07}.
Eventually the energy for the two large scale friction mechanisms
become approximately equal and evolve similarly with time. 
We have also verified that the large scale structures for these two 
cases (not shown) are qualitatively the same. 
The inset of Fig.~\ref{damp.fig} reports the late time ($t > 7.5$)
 evolution
of the  transverse integral length
scale $L_{22,1}$ defined in terms of the two-point
correlation as
(with no sum over Greek subscripts) 
\beq
\label{lint.eq}
L_{\alpha\alpha,\beta}=
\frac{1}{\la u_\alpha^2 \ra}
\int_0^\infty \bla u_\alpha(\ux) 
u_\alpha (\ux+r \mathbf{e}_\beta) \bra \;,
\eeq
along the direction of the unit vector 
$\mathbf{e}_\beta$. The integral scales for the case where 
energy is removed from the thicker shell are smaller than for the case
where energy is removed from the thinner shell and are thus contaminated
by the periodic boundary conditions to a lesser extent.
 
\section{Conclusions}
\label{sec4}
In this study we have used direct numerical simulations
to study the response of rotating turbulence to ``precession-like''
perturbation.
A major emphasis has been to examine the large scale structure
and energy transfer characteristics
when the orientation of the rotation axis is repeatedly
changed.

In the case of uniform solid-body rotation 
with a fixed rotation direction, the
spectral transfer and hence dissipation
is greatly reduced by rotation. 
If the orientation of the rotation axis is changed with a
time scale comparable with the rotation time scale,
the down scale energy transfer and hence dissipation is shown
to increase. After a transient period the kinetic energy reaches
a quasi-stationary state as the energy input by forcing is 
balanced by the dissipation at the small scales. The large scales 
are devoid of columnar structures typically seen in 
rotating flows and resemble that of an isotropic state.
A quantitative assessment of the 
degree of isotropization due to the change in the
orientation of the rotation axis 
will require a systematic projection onto the eigenfunctions of the 
group of rotation and will be reported 
elsewhere \cite{LP05}. 

This work is a first step in studying the influence of precession-like perturbation 
on rotating turbulence. We have neglected the time dependent
precession term $d(\omgv \times \ur)/dt$ under the assumption
that the instantaneous perturbation induced by the sudden change in the 
rotation axis does not affect the long-time dynamical evolution and/or 
the evolution of the fluid region close enough to the rotation axis.
{Using a penalization technique the non-homogeneous precession term 
$d(\omgv \times \ur)/dt$ can be taken into account exactly \cite{kai05}.} 
Another potential source of spurious effects is due to periodic boundary 
conditions, which force the large-scale columns to wrap around the lattice, 
something that would not be possible in presence of a solid boundary \cite{fabien}.
The robustness of these approximations will be quantified in a study presented elsewhere.
 
\section{Acknowledgments}
We acknowledge P. Mininni for useful discussions.
Annick Pouquet is thankful to LASP for its hospitality.
This work was funded by the European Research Council under the 
European Community’s Seventh Framework Program, 
ERC Grant Agreement No.~$339032$.
We acknowledge the CINECA initiatives INF14\_fldturb and 
IscrC\_RotEuler for the availability of 
high performance computing resources and support.

\noindent
All authors contributed equally to the paper. 
\bibliographystyle{epj}
\bibliography{zebib}

\begin{thebibliography}{26}

\bibitem{camb97}
{C. Cambon, N. N. Mansour and F. S. Godeferd}, J. Fluid Mech. \textbf{227}, 303
  (1997)

\bibitem{mans91}
{N. N. Mansour, T. Shih and W. C. Reynolds}, Phys. Fluids \textbf{3}, 2421
  (1991)

\bibitem{SM14}
{B. Saint-Michel, B. Dubrulle, L. Marié, F. Ravelet and F. Daviaud}, New J.
  Phys. \textbf{16}, 063037 (2014)

\bibitem{SM15}
{S. Thalabard, B. Saint-Michel, E. Herbert, F. Daviaud and B. Dubrulle}, New J.
  Phys. \textbf{17}, 063006 (2015)

\bibitem{Pouquet2010}
{A. Pouquet and P. D. Mininni}, Phil. Trans. R. Soc. A \textbf{368}, 1635
  (2010)

\bibitem{sen12}
{A. Sen, P. D. Mininni, D. Rosenberg and A. Pouquet}, Phys. Rev. E \textbf{86},
  036319 (2012)

\bibitem{xia11}
{H. Xia, D. Byrne, G. Falkovich and M. Shats}, Nature Phys. \textbf{7}, 321
  (2011)

\bibitem{kida11}
{S. Kida}, J. Fluid Mech. \textbf{680}, 150 (2011)

\bibitem{goto07}
{S. Goto, N. Ishii, S. Kida, and M. Nishioka}, Phys. Fluids \textbf{19}, 061705
  (2007)

\bibitem{Malkus}
{W. V. R. Malkus}, {Science} \textbf{160}, 259 (1968)

\bibitem{goto14}
{S. Goto, A. Matsunaga, M. Fujiwara, M. Nishioka, S. Kida, M. Yamato, M. and S.
  Tsuda}, Phys. Fluids \textbf{26}, {055107} (2014)

\bibitem{gspan}
{H. P. Greenspan}, \emph{{The Theory of Rotating Fluids}} ({Cambridge
  Monographs on Mech. \& App. Math., Breukelen Press}, 1990)

\bibitem{nore}
{C. Nore, J. L\'eorat, J.-L. Guermond and F. Luddens}, {J. Phys.: Conf. Series}
  \textbf{318}, 072034 (2011)

\bibitem{triana}
{S. A. Triana, D. S. Zimmerman and D. P. Lathrop}, {J. Geophys. Res.}
  \textbf{117}, B04103 (2012)

\bibitem{saw91}
{B. L. Sawford}, Phys. Fluids \textbf{3}, 1577 (1991)

\bibitem{zee94}
{O. Zeman}, Phys. Fluids \textbf{6}, 3221 (1994)

\bibitem{krs95}
K.R. Sreenivasan, Phys. Fluids \textbf{7}, 2778 (1995)

\bibitem{zhou95}
{Y. Zhou}, Phys. Fluids \textbf{7}, 2092 (1995)

\bibitem{min12}
{P. D. Mininni, D. Rosenberg and A. Pouquet}, J. Fluid Mech. \textbf{699}, 263
  (2012)

\bibitem{MY.II}
A.S. Monin, A.M. Yaglom, \emph{Statistical Fluid Mechanics}, Vol.~2 (MIT Press,
  1975)

\bibitem{map09}
P.D. Mininni, A.~Alexakis, A.~Pouquet, Phys. Fluids \textbf{21}(1), 015108
  (2009)

\bibitem{SY93}
{L. M. Smith and V. Yakhot}, Phys. Rev. Lett. \textbf{71}, 352 (1993)

\bibitem{cher07}
{M. Chertkov, C. Connaughton, I. Kolokolov and V. Lebedev}, Phys. Rev. Lett.
  \textbf{99}, 084501 (2007)

\bibitem{LP05}
{L. Biferale and I. Procaccia}, {Phys. Rep.} \textbf{414}, 43 (2005)

\bibitem{kai05}
{K. Schneider}, {Computers \& Fluids} \textbf{34}, 1223 (2005)

\bibitem{fabien}
{F. S. Godeferd and F. Moisy}, {App. Mech. Rev.} \textbf{67}, 030802 (2015)

\end{thebibliography}


@STRING{JFM = {J. Fluid Mech.}}
@STRING{POF = {Phys. Fluids}}
@STRING{PFA = {Phys. Fluids A}}
@STRING{JOT = {J. Turb.}}
@STRING{PRE = {Phys. Rev. E}}
@STRING{ARFM = {Annu. Rev. Fluid Mech.}}
@STRING{PRL = {Phys. Rev. Lett.}}
@STRING{FTC = {Flow Turbul. \& Combust.}}
@STRING{EPL = {Europhys. Lett.}}
@STRING{PTRSA = {Phil. Trans. R. Soc. A}}
@STRING{NJP = {New J. Phys.}}


@ARTICLE{sreeni91,  
	AUTHOR={K. R. Sreenivasan},
	TITLE={Do scalar fluctuations in turbulent shear flows posses local universality?},
	JOURNAL={Physica},      
        volume={51}, 
	PAGES={567-568},
	YEAR={1991}}
@BOOK{Fri95,
	AUTHOR={U. Frisch},       
	TITLE={Turbulence},                                                   
	PUBLISHER={Cambridge University Press},
	YEAR={1995}}
@ARTICLE{EPFOR,
        AUTHOR={V. Eswaran and S. B. Pope},
        TITLE={An Examination of Forcing in Direct Numerical Simulations of Turbulence},
        JOURNAL={Comput. Fluids},
        PAGES={257-278},
  volume = {16},
        YEAR={1988}}
@ARTICLE{K41a,
  AUTHOR={A. N. Kolmogorov},
  TITLE={Local structure of turbulence in an incompressible fluid for very
  large Reynolds numbers},
  JOURNAL={Dokl. Akad. Nauk. SSSR},
  VOLUME={30},
  PAGES={299-303},
  YEAR={1941a}}
@ARTICLE{K41b,
  AUTHOR={A. N. Kolmogorov},
  TITLE={Dissipation of energy in locally isotropic turbulence},
  JOURNAL={Dokl. Akad. Nauk. SSSR},
  VOLUME={434},
  PAGES={16-18},
  YEAR={1941b}}
@ARTICLE{K62,
  AUTHOR={A. N. Kolmogorov},
  TITLE={A refinement of previous hypothesis concerning the
  local structure of turbulence in viscous incompressible fluid
  at high Reynolds number},
  JOURNAL=JFM,
  VOLUME={13},
  PAGES={82-85},
  YEAR={1962}}
@ARTICLE{OP96,
  AUTHOR={M. R. Overholt and S. B. Pope},
  TITLE={Direct numerical simulation of a passive scalar with imposed
  mean gradient in isotropic turbulence},
  JOURNAL=POF,
 volume={8},
  PAGES={3128-3148},
  YEAR={1996}}
@ARTICLE{Breth,
  AUTHOR={G. Brethouwer and  J. C. R. Hunt and F. T. M. Nieuwstadt},
  TITLE={Micro-Structure and Lagrangian statistics of the scalar field
  with a mean gradient in isotropic turbulence},
  JOURNAL={J. Fluid. Mech.},
  PAGES={193-225},  
  YEAR={2003}}
@ARTICLE{fox96,
  AUTHOR={R. 0. Fox},                     
  TITLE={On velocity-conditioned scalar mixing i homogeneous turbulence},
  JOURNAL=POF,
  NUMBER={10}, 
  PAGES={2678-2691},
  YEAR={1996}}
@ARTICLE{pk98, 
  AUTHOR={P. K. Yeung},
  TITLE={Correlations and conditional statistics in differential diffusion: Scalars with uniform mean gradients},
  JOURNAL=POF,
  NUMBER={10},
  PAGES={2621-2635},
  YEAR={1998}}
@ARTICLE{pkxukrs2001,
  AUTHOR={P. K. Yeung and S. Xu and K. R. Sreenivasan},
  TITLE={Schmidt number effects on turbulent transport with uniform mean scalar gradient},                         
  JOURNAL=POF,
  PAGES={4178-4191},
  volume = {{14}},
  YEAR={2002}}
@ARTICLE{ved,
  AUTHOR={P. Vedula and P. K. Yeung and R. O. Fox},                  
  TITLE={Dynamics of scalar dissipation in isotropic turbulence: a numerical andmodelling study},
  JOURNAL=JFM,
  PAGES={29-60},  
 volume               = {433},
  YEAR={2001}}
@ARTICLE{pkxudonz,
  AUTHOR={P. K. Yeung and S. Xu, D.A. Donzis and K. R. Sreenivasan},
  TITLE={Simulations of Three-Dimensional Turbulent Mixing for Schmidt Numbers of Order 1000},
  JOURNAL={Flow Turb. Comust.},
  PAGES={333-347},
  YEAR={2004}}
@ARTICLE{ash,        
  AUTHOR={W. T. Ashurst and A. R. Kerstein and R. M. Kerr and C. H. Gibson},
  TITLE={Alignment of vorticity and scalar gradient with strain rate in simulated Navier-Stokes turbulence},
  JOURNAL=POF,
  volume = {{30}},
  PAGES={2343-2353},
  YEAR={1987}}
@ARTICLE{batch59,
  AUTHOR={G. K. Batchelor},
  TITLE={Small-scale variation
  of convected quantities like temperature in turbulent fluid {P}art 1.
  {G}eneral discussion and the case of small conductivity},
  JOURNAL=JFM,
  volume = {{5}},
  PAGES={113-139},
  DOI = {{10.1017/S002211205900009X}},
  ISSN = {{0022-1120}},
  Unique-ID = {{ISI:A1959WH05100009}},
  YEAR={1959}}
@ARTICLE{War20,
  AUTHOR={Z. Warhaft},
  TITLE={Passive Scalars IN Turbulent Flows},
  JOURNAL=ARFM,
  volume = {{32}},
  PAGES={203--240},
  YEAR={2000}}
@article{corrsin.1951,
 author               = {S. Corrsin},
 journal              = {J. Appl. Phys.},
 pages                = {469-473},
 title                = {On the spectrum of isotropic temperature fluctuation},
 volume               = {22},
 year                 = {1951},
 }
@ARTICLE{obuk1949,
  author = {Obukhov, A. M.},
  title = {The structure of the temperature field in a turbulent flow},
  journal = {Dokl. Akad. Nauk. SSSR},
  year = {1949},
  volume = {39},
  pages = {391},
  owner = {donzis},
  timestamp = {2009.12.02}
}
@ARTICLE{lap2001,
  author = {A. La Porta and G. A. Voth and A. M. Crawford and J. Alexander and E. Bodenschatz},
  title = {Fluid particle accelrations in fully developed turbulence},
  journal = {Nature},
  year = {2001},
  pages = {1017-1019},
}
@CONFERENCE{DYP2008,
  author = {Donzis, D. A. and Yeung, P. K. and Pekurovsky, D.},
  title = {Turbulence simulations on ${O}(10^4)$ processors},
  booktitle = {Proc. {TeraGrid} '08 Conf.},
  year = {2008},
  timestamp = {2008.09.15}
}

@ARTICLE{DYS2008,
  author = {Donzis, D. A. and Yeung, P. K. and Sreenivasan, K. R.},
  title = {Dissipation and enstrophy in isotropic turbulence: Resolution effects and scaling in direct numerical simulations},
  journal = POF,
  year = {2008},
  volume = {20},
  pages = {045108},
}
@ARTICLE{const94,
  author = {P. Constantin},
  title = {Geometric statistics in turbulence},
  journal = {SIAM Review},
  year = {1994},
  pages = {73-98},
}
@ARTICLE{frisch90,
  author = {U. Frisch and S.A. Orszag},
  title = {Turbulence: Challenges for theory and experiment},
  journal = {Physics Today},
  year = {1990},
  pages = {24-32},
}
@ARTICLE{SA97,
  author = {Sreenivasan, K. R. and Antonia, R. A.},
  title = {The phenomenology of small-scale turbulence},
  journal = ARFM,
  year = {1997},
  volume = {29},
  pages = {435-472},
  file = {:pdfs/SA1997.pdf:PDF},
  owner = {donzis},
  timestamp = {2008.06.19}
}
@ARTICLE{BHT,
  author = {Batchelor, G. K. And Howells, I. D. And Townsend, A. A.},
  title = {Small-scale variation of convected quantities like temperature in
        turbulent fluid. Part 2. {T}he case of large conductivity},
  journal = JFM,
  year = {1959},
  volume = {{5}},
  pages = {{134-139}},
}
@ARTICLE{DY10,
  author = {D. A. Donzis and P. K. Yeung},
  title = {Resolution effects and scaling in numerical simulations of
  passive scalar mixing in turbulence},
  journal = {Physica D},
  volume = {{239}},
  year = {2010},
  pages = {1278-1287},
}
@ARTICLE{GAK,
  author = {Gibson, C. H. and Ashurst, W. T. and Kerstein, A. R.},
  title = {Mixing of strongly diffusive passive scalarss like temperature 
          by turbulence},
  journal = {J. Fluid Mech.},
  year = {1988},
  pages = {261-293},
}
@BOOK{grama,
         AUTHOR={A. Grama and A. Gupta and G. Karypis and V. Kumar},
         TITLE={Introduction to Parallel Computing},
         PUBLISHER={Pearson Education Limited},
         YEAR={2003}}
@ARTICLE{moore,
  author = {G. E. Moore},
  title = {Cramming more components onto integrated circuits},
  journal = {Electronics},
  year = {1965},
}
@BOOK{bader,
         AUTHOR={D. A. Bader},
         TITLE={Petascale Computing Algorithms and Applications},
         PUBLISHER={Chapman \& Hall},
         YEAR={2008}}
@BOOK{pope,
         AUTHOR={S. B. Pope},
         TITLE={Turbulent Flows},
         PUBLISHER={Cambridge University Press},
         YEAR={2000},
}
@BOOK{TL,
         AUTHOR={H. Tennekes and J. L. Lumley},
         TITLE={A first course in turbulence},
         PUBLISHER={The {MIT} press},
         YEAR={1972},
}
@BOOK{rich,
         AUTHOR={L. F. Richardson},
         TITLE={Weather Prediction by Numerical Process},
         PUBLISHER={Cambridge University Press},
         YEAR={1922},
}
@book{usingomp,
 author = {Chapman, B. and Jost, G. and Pas, R. V.},
 title = {Using {O}pen{MP}: {P}ortable {S}hared {M}emory {P}arallel {P}rogramming ({S}cientific and {E}ngineering {C}omputation)},
 year = {2007},
 isbn = {0262533022, 9780262533027},
 publisher = {The {MIT} Press},
}
@MISC{openmp,
      TITLE={{O}pen{MP} Application Program Interface Version 3.0},
      YEAR={2008},
      url={http://openmp.org/wp/}}
@MISC{mpidoc,
      TITLE={{MPI}: A Message Passing Interface Standard Version 2.2},
      YEAR={2009},
      PUBLISHER={Message Passing Interface Forum},
      howpublished={http://www.mpi-forum.org/}}
@BOOK{usingmpi,
      AUTHOR={W. Gropp and E. Lusk and R. Thakur},       
      TITLE={Using {MPI}-2},                                                 
      PUBLISHER={Cambridge University Press},
      YEAR={1999}}
@ARTICLE{MM98,
  author = {Moin, P. and Mahesh, K.},
  title = {Direct numerical simulation: A tool in turbulence research},
  journal = ARFM,
  volume = {{30}},
  year = {1998},
  pages = {{539-578}},
  abstract = {{We review the direct numerical simulation (DNS) of turbulent hows.
        We stress that DNS is a research tool, and not a brute-force solution
        to the Navier-Stokes equations for engineering problems. The wide
        range of scales in turbulent flows requires that care be taken in
        their numerical solution. We discuss related numerical issues such
        as boundary conditions and spatial and temporal discretization. Significant
        insight into turbulence physics has been gained from DNS of certain
        idealized flows that cannot be easily attained in the laboratory.
        We discuss some examples. Further, we illustrate the complementary
        nature of experiments and computations in turbulence research. Examples
        are provided where DNS data has been used to evaluate measurement
        accuracy. Finally, we consider how DNS has impacted turbulence modeling
        and provided further insight into the structure of turbulent boundary
        layers.}},
  file = {:pdfs/MM98.pdf:PDF},
  unique-id = {{ISI:000071880700019}}
}
@ARTICLE{EP88,
  author = {Eswaran, V. and Pope, S. B.},
  title = {An examination of forcing in direct numerical simulations of turbulence},
  journal = {Comput. Fluids},
  year = {1988},
  volume = {16},
  pages = {257-278},
  owner = {donzis},
  timestamp = {2008.06.19}
}
@article{rogallo,
  author = {Rogallo, R. S.},
  title = {Numerical experiments in homogeneous turbulence},
  institution = {NASA},
  journal = {NASA Tech.~{}Memo 81315, NASA Ames Research Center},
  year = {1981},
  publisher = {NASA Ames Research Center, Moffett Field, CA.},
}
@article{rogallo77,
  author = {Rogallo, R. S.},
  title = {An {ILLIAC} Program for the Numerical Simulation of Homogeneous
Incompressible Turbulence},
  institution = {NASA},
  journal = {NASA Tech.~{}Memo},
  year = {1977},
  publisher = {NASA Ames Research Center, Moffett Field, CA.},
}
@ARTICLE{YS2005,
  author = {Yakhot, V. and Sreenivasan, K. R.},
  title = {Anomalous scaling of structure functions and dynamic constraints
        on turbulence simulations},
  journal = {J. Stat. Phys.},
  year = {2005},
  volume = {121},
  pages = {823-841},
  file = {:pdfs/YS2005.pdf:PDF},
  owner = {donzis},
  timestamp = {2008.06.19}
}
@ARTICLE{fftw,
  author = {M. Frigo and S. G. Johnson},
  title = {The Design and Implementation of {FFTW3}},
  journal = {Proc. IEEE},
  year = {2005},
  pages = {216-231},
}
@Article{FFTW05,
  author = 	 {Frigo, M. and Johnson, S. G.},
  title = 	 {The Design and Implementation of {FFTW3}},
  journal = 	 {Proc.  IEEE},
  year = 	 2005,
  pages =	 {216--231},
  note =	 {Special issue on ``Program Generation, Optimization, and Platform Adaptation''}
}
@MISC{bwaters,
      title = {Blue Waters},
      howpublished={http://www.ncsa.illinois.edu/BlueWaters/}}
@MISC{p3dfft,
      author = {Pekurovsky, D.},
      title = {{P3DFFT}},
      howpublished={http://www.sdsc.edu/us/resources/p3dfft/index.php},
      year = {2008}}
@misc{BW,
      title  = {{Blue Waters}},
      howpublished   = "\url{http://www.ncsa.illinois.edu/BlueWaters/}",
      year = {2012},
    }
@ARTICLE{warhaft2009,
  author = {Z.  Warhaft},
  title = {Why we need experiments at high {R}eynolds numbers},
  journal = {Fluid Dyn. Res.},
  volume = {41},
  year = {2009},
  pages = {021401},
  abstract = {This brief review focuses on the effect of Reynolds number on turbulence
	small-scale anisotropy and intermittency, on Lagrangian experiments
	of fluid and inertial particles and on complex turbulent flows. Although
	Taylor scale Reynolds numbers of 103 (equivalent to turbulence Reynolds
	number based on the integral scale of 105) can be achieved in the
	laboratory, it is argued that there is still a need to do experiments
	(and computation) at even higher Reynolds numbers to resolve outstanding
	basic and applied issues in turbulence.},
  file = {:warhaft2009.pdf:PDF},
  owner = {diego},
  timestamp = {2009.10.23},
  url = {http://stacks.iop.org/1873-7005/41/021401}
}
@ARTICLE{IGK2009,
  author = {Ishihara, T. and Gotoh, T. and Kaneda, Y.},
  title = {Study of high-{R}eynolds number isotropic turbulence by direct numerical
	simulation},
  journal = ARFM,
  volume = {41},
  year = {2009},
  pages = {165--180},
  abstract = {We review studies of the statistics of isotropic turbulence in an
	incompressible fluid at high Reynolds numbers using direct numerical
	simulation (DNS) from the viewpoint of fundamental physics. The Reynolds
	number achieved by the largest DNS, with 4096(3) grid points, is
	comparable with the largest Reynolds number in laboratory experiments.
	The high-quality DNS data in the inertial subrange and the dissipative
	range enable the examination of detailed statistics at small scales,
	such as the normalized energy-dissipation rate, energy and energy-flux
	spectra, the intermittency of the velocity gradients and increments,
	scaling exponents, and flow-field structure. We emphasize basic questions
	of turbulence, universality in the sense of Kolmogorov's theory,
	and the dependence of the statistics on the Reynolds number and scale.},
  af = {Ishihara, TakashiEOLEOLGotoh, ToshiyukiEOLEOLKaneda, Yukio},
  di = {10.1146/annurev.fluid.010908.165203},
  file = {:pdfs/IGK2009.pdf:PDF},
  owner = {donzis},
  sn = {0066-4189},
  timestamp = {2009.09.29},
  ut = {ISI:000262972800010}
}
@ARTICLE{gibson1968b,
  author = {Gibson, C. H.},
  title = {Fine Structure of Scalar Fields Mixed By Turbulence. {II}. {S}pectral
	Theory},
  journal = {Phys. Fluids},
  year = {1968},
  pages = {2316--2327},
  file = {:pdfs/gibson1968b.pdf:PDF},
  owner = {donzis},
  sn = {1070-6631},
  tc = {70},
  timestamp = {2009.11.30},
  ut = {WOS:A1968C165400003}
}
@ARTICLE{YDS2012,
  author = {Yeung, P. K. and Donzis, D. A. and Sreenivasan, K. R.},
  title = {Dissipation, enstrophy and pressure statistics in turbulence
simulations at high Reynolds numbers},
  journal = JFM,
  year = {2012},
  volume = {{700}},
  pages = {5-15},
}
@ARTICLE{a2aopt08,
  author = {Kumar, S.  and Sabharwal, Y and Heidelberger, P.},
  title = {Optimization of All-to-all communication on the
Blue Gene/L Supercomputer},
  journal = {International Conference on Parallel Processing},
  year = {2008},
  pages = {320-329},
}
@BOOK{canuto,
         AUTHOR={C. Canuto and M. Y. Hussaini and A. Quarteroni and T. A. Zang},
         TITLE={Spectral Methods in Fluid Dynamics},
         PUBLISHER={Springer-Verlag},
         YEAR={1987},
}
@ARTICLE{PO71,
  AUTHOR={G. S. Patterson and S. A. Orszag},
  TITLE={Spectral Calculations of Isotropic Turbulence: Efficient Removal
  of Aliasing Interactions},
  JOURNAL=POF,
  PAGES={2538-2541},
  YEAR={1971}}
@ARTICLE{yp89,
  author = {Yeung, P. K. and Pope, S. B.},
  title = {Lagrangian statistics from direct numerical simulations of isotropic
	turbulence},
  journal = {J. Fluid Mech.},
  year = {1989},
  volume = {{207}},
  pages = {{531-586}},
  file = {:pdfs/yp89.pdf:PDF},
  unique-id = {{ISI:A1989AZ88800022}},
}
@ARTICLE{GT07,
  author = {Gropp, W. and Thakur, R.},
  title = {Thread-safety in {MPI} implementation: Requirement and analysis},
  journal = {Parallel Computing},
  year = {2006},
  pages = {{595-604}},
}
@ARTICLE{mpimill,
  author = {Balaji, P. and Buntinas,D. and Goodell, D. and Gropp, W. and
           Kumar, S. and Lusk, E. and Thakur, R. and Traff, J. L.},
  title = {{MPI} on a Million Processors},
  journal = {Lecture Notes in Computer Science},
  year = {2009},
  pages = {{20-30}},
}
@ARTICLE{ompmem,
  author = {Hoeflinger, J.P. and de Supinski, B.R.},
  title = {The {O}pen{MP} memory model},
  journal = {Lecture Notes in Computer Science},
  year = {2008},
  pages = {{167-177}},
}
@inproceedings{ovrlap,
 author = {Doi, J. and Negishi, Y.},
 title = {Overlapping Methods of All-to-All Communication and {FFT} Algorithms 
for Torus-Connected Massively Parallel Supercomputers},
 booktitle = {Proc.~{} of the 2010 ACM/IEEE Int.~{} Conf. for High 
Performance Computing, Networking, Storage and Analysis},
 series = {{SC} '10},
 year = {2010},
 isbn = {978-1-4244-7559-9},
 number = {},
 pages = {1--9},
 numpages = {9},
 url = {http://dx.doi.org/10.1109/SC.2010.38},
 doi = {10.1109/SC.2010.38},
 acmid = {1884652},
 publisher = {IEEE Computer Society},
 address = {Washington, DC, USA},
}
@inproceedings{ whyhybrid,
Author = {Tsuji, M. and Sato, M.},
Title = {{Performance evaluation of OpenMP and MPI hybrid programs on a large
   scale multi-core multi-socket cluster, T2K Open Supercomputer}},
Booktitle = {{{I}nt.~{} Conf.~{} on Parallel Process. Workshops (ICPPW
   2009)}},
Year = {{2009}},
Pages = {{206-213}},
Abstract = {{Non-uniform memory access (NUMA) systems, where each processor has its
   own memory, have been popular platform in high-end computing. While some
   early studies had reported that a flat-MPI programming model
   outperformed an OpenMP/MPI hybrid programming model on SMP clusters, the
   hybrid of a shared-memory, thread-based programming and a
   distributed-memory, message passing programming is considered to be a
   promising programming model on the multi-core multi-socket NUMA
   clusters. We explore the performance of the OpenMP/MPI hybrid
   programming model on a large scale multi-core multi-socket cluster
   called T2K Open Supercomputer. Both of benchmark (NPB, NAS Parallel
   Benchmarks) and application (RSDFT, Real-Space Density Function Theory)
   codes are considered. The hybridization for the RSDFT code is also
   shown. Our experiments show that the multi-core multi-socket cluster can
   take advantage of the hybrid programming model when it uses MPI across
   sockets and OpenMP within sockets.}},
Publisher = {{IEEE}},
Type = {{Proceedings Paper}},
Language = {{English}},
Affiliation = {{Tsuji, M (Reprint Author), Univ Tsukuba, Ctr Computat Sci, Tsukuba, Ibaraki, Japan..
   Tsuji, Miwako; Sato, Mitsuhisa, Univ Tsukuba, Ctr Computat Sci, Tsukuba, Ibaraki, Japan.}},
DOI = {{10.1109/ICPPW.2009.73}},
ISBN = {{978-1-4244-4923-1}},
Keywords = {{performance evaluation; OpenMP; MPI}},
Research-Areas = {{Computer Science; Engineering}},
Web-of-Science-Categories  = {{Computer Science, Theory \& Methods; Engineering, Electrical \&
   Electronic}},
Author-Email = {{tsuji@hpcs.cs.tsukuba.ac.jp
   msato@cs.tsukuba.ac.jp}},
Number-of-Cited-References = {{6}},
Times-Cited = {{0}},
Doc-Delivery-Number = {{BUO22}},
Unique-ID = {{ISI:000289915300029}},
}
@article{ scan,
Author = {Blelloch, G. E.},
Title = {{Scans as Primitive Parallel Operations}},
Journal = {{IEEE Transcations on Computers}},
Year = {{1989}},
Pages = {{1526-1538}},
Publisher = {{IEEE Computer Soc}},
Address = {{10662 Los Vaqueros Circle, PO BOX 3014, Los Alamitos, CA 90720-1264}},
Type = {{Article}},
Language = {{English}},
Affiliation = {{Blelloch, G. E. (Reprint Author), Carnegie Mellon UNIV,SCH COMP SCI,Pittsburgh,PA 15213, USA..}},
DOI = {{10.1109/12.42122}},
ISSN = {{0018-9340}},
Research-Areas = {{Computer Science; Engineering}},
Web-of-Science-Categories  = {{Computer Science, Hardware \& Architecture; Engineering, Electrical \&
   Electronic}},
Number-of-Cited-References = {{48}},
Times-Cited = {{106}},
Journal-ISO = {{IEEE Trans. Comput.}},
Doc-Delivery-Number = {{AW991}},
Unique-ID = {{ISI:A1989AW99100005}},
}
@article{ donz05,
Author = {Donzis, D. A. and Sreenivasan, K. R. and Yeung, P. K.},
Title = {{Scalar dissipation rate and dissipative anomaly in isotropic turbulence}},
Journal = JFM,
Year = {{2005}},
  volume = {{532}},
Pages = {{199-216}},
Abstract = {{We examine available data from experiment and recent numerical
   simulations to explore the supposition that the scalar dissipation rate
   in turbulence becomes independent of the fluid viscosity when the
   viscosity is small and of scalar diffusivity when the diffusivity is
   small. The data are interpreted in the context of semi-empirical
   spectral theory of Obukhov and Corrsin when the Schmidt number, Sc, is
   below unity, and of Batchelor's theory when Sc is above unity. Practical
   limits in terms of the Taylor-microscale Reynolds number, R-lambda, as
   well as Sc, are deduced for scalar dissipation to become sensibly
   independent of molecular properties. In particular, we show that such an
   asymptotic state is reached if RlambdaSc1/2 >> 1 for Sc < 1, and if
   ln(Sc)/R-lambda << 1 for Sc > 1.}},
DOI = {{10.1017/S0022112005004039}},
ISSN = {{0022-1120}},
Unique-ID = {{ISI:000230501600008}},
}
@article{yag49,
Author = {Yaglom, A. M.},
Title = {{{O}n the local structure of a temperature field in a 
turbulent flow}},
Journal = {{{D}okl. {A}kad. {N}auk {NAUK} {SSSR}}},
Year = {{1949}},
 volume               = {69},
Pages = {{743-746}},
ISSN = {{0002-3264}},
Unique-ID = {{ISI:A1949UG33300009}},
}
@article{ kerr85,
Author = {Kerr, R. M.},
Title = {{Higher-order derivative correlations and the alignment of small-scale
   structures in isotropic numerical turbulence}},
Journal = JFM,
Year = {{1985}},
  volume = {{153}},
Pages = {{31-58}},
DOI = {{10.1017/S0022112085001136}},
ISSN = {{0022-1120}},
Unique-ID = {{ISI:A1985AHV9200003}},
}
@ARTICLE{dissana,
author={Sreenivasan, K. R.},
title={{An update on the energy dissipation rate in isotropic turbulence}},
journal=POF,
volume={10},
year={1998},
pages={528-529}
}
@article{ DSY10,
Author = {Donzis, D. A. and Sreenivasan, K. R. and Yeung, P. K.},
Title = {{The Batchelor Spectrum for Mixing of Passive Scalars in Isotropic
   Turbulence}},
Journal = FTC,
Year = {{2010}},
  volume = {{85}},
Pages = {{549-566}},
Abstract = {{We examine the support for the Batchelor spectrum from well-resolved
   simulations of high-Schmidt-number mixing in isotropic turbulence, and
   resolve a conundrum with respect to the numerical value of the
   prefactor, also known as the Batchelor constant. Our conclusion is that
   the most probable value of the most compressive principal strain rate is
   more relevant than its mean, at least asymptotically.}},
DOI = {{10.1007/s10494-010-9271-6}},
ISSN = {{1386-6184}},
Unique-ID = {{ISI:000283943000015}},
}
@article{ sreenitauv,
Author = {Sreenivasan, K. R. and Tavoularis, S.},
Title = {{On the skewness of the temperature derivative in turbulent flows}},
Journal = JFM,
Year = {{1980}},
  volume = {101},
Pages = {{783-795}},
DOI = {{10.1017/S0022112080001929}},
ISSN = {{0022-1120}},
Unique-ID = {{ISI:A1980LD74800007}},
}
@article{ anselmet,
Author = {Anselmet, F. and Gagne, Y. and Hopfinger, E. J. and Antonia, R. A.},
Title = {{High-order velocity structure functions in turbulent shear flows}},
Journal = JFM,
Year = {{1984}},
Pages = {{63-89}},
DOI = {{10.1017/S0022112084000513}},
ISSN = {{0022-1120}},
Unique-ID = {{ISI:A1984SN69100005}},
}
@ARTICLE{obukhov,
  AUTHOR = {Obukhov, A.},
  TITLE = {Some specific features of atmospheric turbulence},
  JOURNAL = JFM,
  VOLUME = {13},
  YEAR = {{1962}},
  PAGES = {{77-81}},
}
@ARTICLE{krsh,
AUTHOR={Stolovitzky, G. and Kailasnath, P. and Sreenivasan, K. R.},
TITLE={Kolmogorov's refined similarity hypotheses},
JOURNAL=PRL,
YEAR={1992},
VOLUME={69},
PAGES={1178-1181}}
@article{ krshp,
Author = {Stolovitzky, G. and Kailasnath, P. and Sreenivasan, K. R.},
Title = {{Refined similarity hypotheses for passive scalars mixed by turbulence}},
Journal = JFM,
volume = {297},
Year = {{1995}},
Pages = {{275-291}},
DOI = {{10.1017/S0022112095003090}},
ISSN = {{0022-1120}},
Unique-ID = {{ISI:A1995RT40200013}},
}
@ARTICLE{menkrs87,
AUTHOR={Meneveau, C. and Sreenivasan, K. R.},
TITLE={{Simple Multifractal Cascade Model for Fully Developed Turbulence}},
JOURNAL=PRL,
YEAR={1987},
VOLUME={59},
PAGES={1424-1427}}
@ARTICLE{krsht,
AUTHOR={Chen, S. and Sreenivasan, K. R. and Nelkin, M. and Cao, N.},
TITLE={{Refined Similarity Hypothesis for Transverse Structure Functions
       in Fluid Turbulence}},
JOURNAL=PRL,
YEAR={1997},
VOLUME={79},
PAGES={2253-2256}}
@ARTICLE{ wangkrsh,
AUTHOR={Wang, L. P. and Chen, S. and Brasseur, J. G. and Wyngaard, J. C.},
Title={{Examination of hypotheses in the Kolmogorov refined turbulence theory
   through high-resolution simulations. {P}art 1. {V}elocity field}},
JOURNAL=JFM,
Year={1996},
VOLUME={309},
PAGES={113-156}
}
@article{ wangkrshp,
Author = {Wang, L. P. and Chen, S. and Brasseur, J. G.},
Title = {{Examination of hypotheses in the Kolmogorov refined turbulence theory
   through high-resolution simulations. {P}art 2. {P}assive scalar field}},
Journal = JFM,
  volume = {{400}},
Year = {{1999}},
Pages = {{163-197}},
Abstract = {{Using direct numerical simulations (DNS) and large-eddy simulations
   (LES) of velocity and passive scalar in isotropic turbulence (up to
   512(3) grid points), we examine directly and quantitatively the refined
   similarity hypotheses as applied to passive scalar fields (RSHP) with
   Prandtl number of order one. Unlike previous experimental
   investigations, exact energy and scalar dissipation rates were used and
   scaling exponents were quantified as a function of local Reynolds
   number. We first demonstrate that the forced DNS and LES scalar fields
   exhibit realistic inertial-range dynamics and that the statistical
   characteristics compare well with other numerical, theoretical and
   experimental studies. The Obukhov-Corrsin constant for the k(-5/3)
   scalar variance spectrum obtained from the 512(3) mesh is 0.87 +/- 0.10.
   Various statistics indicated that the scalar field is more intermittent
   than the velocity field. The joint probability distribution of
   locally-averaged energy dissipation epsilon(r) and scalar dissipation
   chi(r) is close to log-normal with a correlation coefficient of 0.25 +/-
   0.01 between the logarithmic dissipations in the inertial subrange. The
   intermittency parameter for scalar dissipation is estimated to be in the
   range 0.43 similar to 0.77, based on direct calculations of the variance
   of ln chi(r). The scaling exponents of the conditional scalar increment
   <(delta(r)theta\textbackslash{}(chi r,epsilon r))over bar> suggest a
   tendency to follow RSHP. Most significantly, the scaling exponent of
   <(delta(r)theta\textbackslash{}(chi r,epsilon r))over bar> over
   epsilon(r) was shown to be approximately -1/6 in the inertial subrange,
   confirming a dynamical aspect of RSHP. In agreement with recent
   experimental results (Zhu et al. 1995; Stolovitzky et al. 1995), the
   probability distributions of the random variable beta(s) =
   delta(r)theta\textbackslash{}(chi r,epsilon r)/(chi(r)(1/2)
   epsilon(r)(-1/6) r(1/3)) were found to be nearly Gaussian. However,
   contrary to the experimental results, we find that the moments of
   beta(s) are almost identical to those for the velocity field found in
   Part 1 of this study (Wang et al. 1996) and are insensitive to Reynolds
   number, large-scale forcing, and subgrid modelling.}},
DOI = {{10.1017/S0022112099006448}},
ISSN = {{0022-1120}},
Unique-ID = {{ISI:000084223600006}},
}
@article{ krskailas,
Author = {Sreenivasan, K. R. and Kailasnath, P.},
Title = {{An update on the intermittency exponent in turbulence}},
Journal = POF,
Year = {{1993}},
  volume = {{5}},
Pages = {{512-514}},
DOI = {{10.1063/1.858877}},
ISSN = {{0899-8213}},
Unique-ID = {{ISI:A1993KJ51600027}},
}
@article{krsad,
Author={Sreenivasan, K. R. and Antonia, R. A. and Danh, H. Q.},
Title={{T}emperature dissipation fluctuations in a turbulent boundary layer},
Journal=POF,
Year={1977},
Volume={20},
Pages={1238-1249}
}
@article{ menevkrs,
Author = {Meneveau, C. and {S}reenivasan, K. R.},
Title = {{The multifractal nature of turbulent energy dissipation}},
Journal = JFM,
Year = {{1991}},
Pages = {{429-484}},
DOI = {{10.1017/S0022112091001830}},
ISSN = {{0022-1120}},
Unique-ID = {{ISI:A1991FG99100022}},
}
@article{ YS2013,
Author = {Yeung, P. K. and Sreenivasan, K. R.},
Title = {{Spectrum of passive scalars of high molecular diffusivity in turbulent
   mixing}},
Journal = JFM,
Year = {{2013}},
volume = {716},
Pages = {{R14}},
Month = {{FEB}},
Abstract = {{We consider the mixing of passive scalars transported in turbulent flow,
   with a molecular diffusivity that is large compared to the kinematic
   viscosity of the fluid. This particular case of mixing has not received
   much attention in experiment or simulation even though the first
   putative theory, due to Batchelor, Howells \& Townsend (J. Fluid Mech.,
   vol. 5, 1959, pp. 134-139), is now more than 50 years old. We study the
   problem using direct numerical simulation of decaying scalar fields in
   steadily sustained homogeneous turbulence as the Schmidt number (the
   ratio of the kinematic viscosity of the fluid to the molecular
   diffusivity of the scalar) is allowed to vary from 1/8 to 1/2048 for two
   values of the microscale Reynolds number, R-lambda approximate to 140
   and approximate to 240. The simulations show that the passive scalar
   spectrum assumes a slope of -17/3 in a range of scales, as predicted by
   the theory, when the Schmidt number is small and the Reynolds number is
   simultaneously large. The observed agreement between theory and
   simulation in the prefactor in the spectrum is not perfect. We assess
   the reasons for this discrepancy by a careful examination of the scalar
   evolution equation in the light of the assumptions of the theory, and
   conclude that the finite range of scales resolved in simulations is the
   main reason. Numerical issues specific to the regime of very low Schmidt
   numbers are also addressed briefly.}},
DOI = {{10.1017/jfm.2012.632}},
Article-Number = {{R14}},
ISSN = {{0022-1120}},
Unique-ID = {{ISI:000314422800014}},
}
@article{YS2014,
Author = {Yeung, P. K. and Sreenivasan, K. R.},
Title = {{Direct numerical simulation of turbulent mixing at very low Schmidt number with a uniform mean gradient}},
Journal = POF,
pages  = {015107},
Year = {2014},
Volume = {26},
}
@INCOLLECTION{GY.2013,
 author = {Gotoh, T. and Yeung, P. K.},
 title = {{Passive scalar transport in turbulence: A computational perpsective}},
 booktitle = {Ten Chapters in Turbulence},
 editor = {P. A. Davidson and Y. Kaneda and K. R. Sreenivasan},
 publisher = {Cambridge University Press},
 year = {2013},
}

@article{OA2002,
 author               = {Orlandi, P. and Antonia, R. A.},
 journal              = JFM,
 pages                = {99--108},
 title                = {Dependence of the nonstationary form of {Y}aglom's equation on the Schmidt number},
 year                 = {2002},
 }

@article{kerr90,
 author               = {Kerr, R. M.},
 journal              = JFM,
 pages                = {309-332},
 title                = {Velocity, scalar and transfer spectra in numerical turbulence},
 year                 = {1990},
 }


@article{ sreeni84,
Author = {Sreenivasan, K. R.},
Title = {{On the scaling of the turbulence energy-dissipation rate}},
Journal = POF,
Year = {{1984}},
Pages = {{1048-1051}},
DOI = {{10.1063/1.864731}},
ISSN = {{1070-6631}},
Unique-ID = {{ISI:A1984SS96600003}},
}
@article{osb84,
Author = {Jackson, D. and Launder, B.},
Title = {{Osborne Reynolds and the publication of his papers on turbulent flow}},
Journal = ARFM,
  volume = {{39}},
Year = {{2007}},
Pages = {{19-35}},
Abstract = {{Following The Royal Society's decision to allow the release of hitherto
   confidential documents concerning referees' reports and other material
   relating to the publication of historical manuscripts, the present paper
   examines the exchanges preceding the publication of the two papers on
   turbulent flow by Osborne Reynolds that have so greatly influenced the
   development of Engineering Fluid Mechanics over the past century. The
   documents cited reveal that, although the earlier experimental paper was
   warmly welcomed, the referees were critical of the subsequent analytical
   contribution. It appears that the publication of the latter paper was
   due mainly to the considerable standing Reynolds had by then acquired,
   in part from the impact of his experimental paper published 12 years
   earlier. The paper also provides a summary of Reynolds' career and
   research prior to his embarking on the research published in the two
   seminal papers.}},
DOI = {{10.1146/annurev.fluid.39.050905.110241}},
ISSN = {{0066-4189}},
ISBN = {{978-0-8243-0739-4}},
Unique-ID = {{ISI:000243900500003}},
}
@article{Taylor38,
author = {Taylor, G. I.}, 
title = {The Spectrum of Turbulence},
volume = {164},
pages = {476-490}, 
year = {1938}, 
doi = {10.1098/rspa.1938.0032}, 
URL = {http://rspa.royalsocietypublishing.org/content/164/919/476.short}, 
eprint = {http://rspa.royalsocietypublishing.org/content/164/919/476.full.pdf+html}, 
journal = {Proc. of the Royal Society of London. Series A - Mathematical and Physical Sciences} 
}
@article{Taylor35,
author = {Taylor, G. I.}, 
title = {Statistical theory of Turbulence},
volume = {151},
pages = {421-444}, 
year = {1935}, 
journal = {Proc. of the Royal Society of London. Series A - Mathematical and Physical Sciences} 
}
@article {rich22,
author = {Richardson, L. F.}, 
title = {Weather prediction by numerical process. Cambridge (University Press), 1922},
journal = {Quarterly Journal of the Royal Meteorological Society},
number = {203},
publisher = {John Wiley & Sons, Ltd},
issn = {1477-870X},
url = {http://dx.doi.org/10.1002/qj.49704820311},
doi = {10.1002/qj.49704820311},
pages = {282--284},
year = {1922},
}
@article{ grant62,
Author = {Grant, H. L. and Stewart, R. W. and Moilliet, A.},
Title = {{Turbulence spectra from a tidal channel}},
Journal = JFM,
Year = {{1962}},
  volume = {{12}},
Pages = {{241-268}},
DOI = {{10.1017/S002211206200018X}},
ISSN = {{0022-1120}},
Unique-ID = {{ISI:A1962WP42900006}},
}
@article{ novi71,
Author = {Novikov,  E. A.},
Title = {{Intermittency and scale similarity in structure of a turbulent-flow}},
Journal = {{Journal of applied mathematics and mechanics-USSR}},
Year = {{1971}},
Pages = {{231-\&}},
DOI = {{10.1016/0021-8928(71)90029-3}},
Unique-ID = {{ISI:A1971L489200008}},
}
@article{ siggia81,
Author = {Siggia, E. D.},
Title = {{Numerical study of small-scale intermittency in 3-dimensional turbulence}},
Journal = JFM,
Year = {{1981}},
Pages = {{375-406}},
DOI = {{10.1017/S002211208100181X}},
ISSN = {{0022-1120}},
Unique-ID = {{ISI:A1981LY59100016}},
}
@article{ kailas92,
Author = {Kailasnath, P. and Sreenivasan, K. R. and Stolovitzky, G.},
Title = {{Probability density of velocity increments in turbulent flows}},
Journal = PRL,
Year = {{1992}},
Pages = {{2766-2769}},
Month = {{MAY 4}},
Abstract = {{Measurements have been made of the probability density function (PDF) of
   velocity increments DELTA-u(r) for a wide range of separation distances
   r. Stretched exponentials provide good working approximations to the
   tails of the PDF. The stretching exponent varies monotonically from 0.5
   for r in the dissipation range to 2 for r in the integral scale range.
   Theoretical forms based on multifractal notions of turbulence agree well
   with the measured PDFs. When the largest scales in the velocity u are
   filtered out, the PDF of DELTA-u(r) becomes symmetric and, for large r,
   close to exponential.}},
DOI = {{10.1103/PhysRevLett.68.2766}},
ISSN = {{0031-9007}},
Unique-ID = {{ISI:A1992HR18300014}},
}
@article{ tabeling96,
Author = {Tabeling, P. and Zocchi, G. and Belin, F. and Maurer, J. and Willaime, H.},
Title = {{Probability density functions, skewness, and flatness in large Reynolds
   number turbulence}},
Journal = PRE,
Year = {{1996}},
Pages = {{1613-1621}},
Month = {{FEB}},
Abstract = {{Probability density functions (PDF) of longitudinal velocity increments
   are measured in an experiment using low temperature helium gas, in which
   a large domain of microscale Reynolds numbers R(lambda) (from 150 to
   5040) is explored. The technique of measurement, which is essentially
   hot wire anemometry, but operating in nonstandard conditions, is
   described. In the inertial range of scales, and for a given scale of
   separation, the PDF are found to be independent of the Reynolds number
   at large R(lambda). Measurements of the skewness and the flatness of the
   velocity derivatives are performed in a range of R(lambda) lying between
   150 and 3200. The surprising result that we find is that these
   quantities first increase with the Reynolds number up to R(lambda)
   approximate to 700 and then cease to increase further, indicating a
   transitional behavior.}},
DOI = {{10.1103/PhysRevE.53.1613}},
ISSN = {{1539-3755}},
Unique-ID = {{ISI:A1996TW97900044}},
}
@article{ shraimansiggia,
Author = {Shraiman, B and Siggia, E},
Title = {{Scalar turbulence}},
Journal = {{Nature}},
Year = {{2000}},
  volume = {405},
Pages = {{639-646}},
Abstract = {{The advection of a passive substance by a turbulent flow is important in
   many natural and engineering settings. The concentration of such a
   substance can exhibit complex dynamic behaviour that shows many
   phenomenological parallels with the behaviour of the turbulent velocity
   field. Yet the statistical properties of this so-called `passive scalar'
   turbulence are decoupled from those of the underlying velocity field.
   Passive scalar turbulence has recently yielded to mathematical analysis,
   and such progress may ultimately lead to a better understanding of the
   still intractable problem of fluid turbulence itself.}},
DOI = {{10.1038/35015000}},
ISSN = {{0028-0836}},
Unique-ID = {{ISI:000087465800036}},
}
@article{ prasad88,
Author = {Prasad, R. R. and Meneveau, C. and Sreenivasan, K. R.},
Title = {{Multifractal nature of the dissipation field of passive scalars in fully 
   turbulent flows}},
Journal = PRL,
  volume = {61},
Year = {{1988}},
Pages = {{74-77}},
DOI = {{10.1103/PhysRevLett.61.74}},
ISSN = {{0031-9007}},
ResearcherID-Numbers = {{Meneveau, Charles/A-3319-2010}},
Unique-ID = {{ISI:A1988P028900019}},
}
@article{ chen98,
Author = {Chen, SY and Kraichnan, RH},
Title = {{Simulations of a randomly advected passive scalar field}},
Journal = {{PHYSICS OF FLUIDS}},
Year = {{1998}},
Pages = {{2867-2884}},
Month = {{NOV}},
Abstract = {{The advection of a passive scalar field by a rapidly decorrelating
   random velocity field with power-law scaling is computed by simulations
   in a cyclic square at resolutions of 4096(2) and 8192(2) grid points.
   Structure functions of the scalar field are measured and inertial-range
   scaling exponents are determined. The conditional mean of the
   scalar-field dissipation term and its moments are found. The results are
   compared with theoretical predictions and with other recent simulations.
   (C) 1998 American Institute of Physics. {[}S1070-6631(98)00211-6]}},
DOI = {{10.1063/1.869808}},
ISSN = {{1070-6631}},
ResearcherID-Numbers = {{Chen, Shiyi/A-3234-2010}},
Unique-ID = {{ISI:000076587500016}},
}
@article{ holzer94,
Author = {Holzer, M. and Siggia, E. D.},
Title = {{Turbulent mixing of a passive scalar}},
Journal = POF,
  volume = {{6}},
Year = {{1994}},
Pages = {{1820-1837}},
Month = {{MAY}},
Abstract = {{The statistically stationary state of a turbulently advected passive
   scalar is studied, with an imposed linear mean gradient in two
   dimensions, via a number of numerical experiments. For a synthetic
   Gaussian velocity field, which is generated by a linear stochastic
   process, and whose spectra and Eulerian correlation time follow
   Kolmogorov scaling on all scales, the exponents of the scalar spectra
   are consistent with 5/3 or 17/3 depending on the diffusivity. For large
   Peclet numbers (Pe), the probability density function (PDF) of the
   scalar gradients perpendicular to the mean is well fit, from about
   0.1-10 times the root-mean-square value, by a stretched exponential with
   exponent -0.6. The PDF for gradients parallel to the mean has similar
   tails and a O(1) skewness for all Pe studied. The scalar has a
   ramp-and-cliff structure similar to that first seen in shear-flow
   experiments with scalars. A physical picture of the mechanism by which
   the ramp-and-cliff features form is given. A second model with the
   velocity evolving under the Euler equations restricted to a band of wave
   numbers produces the k-1 Batchelor spectrum when the scalar is
   dissipated with a hyperdiffusivity (is-proportional-to k4). For physical
   dissipation (is-proportional-to k2), the PDF of the scalar has
   exponential tails, and for gradients less than the cutoff set by the
   maximum strain, the PDF of the gradients is similar to that obtained
   with the stochastic velocity model. The PDF of the dissipation is
   approximately stretched exponential like the gradient PDFs and not
   lognormal. The skewness of the gradients parallel to the mean decreases
   with decreasing autocorrelation time of the velocity, and the gradient
   PDFs assume a limiting form in the white-noise limit.}},
DOI = {{10.1063/1.868243}},
ISSN = {{1070-6631}},
Unique-ID = {{ISI:A1994NH62400019}},
}
@article{ kraich94,
Author = {Kraichnan, R. H.},
Title = {{Anomalous scaling of a randomly advected passive scalar}},
Journal = PRL,
Year = {{1994}},
  volume = {{72}},
Pages = {{1016-1019}},
Month = {{FEB 14}},
Abstract = {{Solenoidal advection in d dimensions may induce inertial-range behavior
   {[}\textbackslash{}T(x, t) - T(x + r, t)\textbackslash{}(n)]
   proportional to r(zeta n(T)) of a passive scalar field T(x, t). If the
   velocity field changes very rapidly in time, the zeta(n)(T) are
   determined by the two-particle eddy-diffusion coefficient eta(r)
   proportional to r(zeta(eta)). An approximation to the
   molecular-diffusion terms yields zeta(n)(T) = 1/2 root
   2nd{[}2-zeta(eta)] + alpha(2) - 1/2 alpha where alpha = d + zeta(eta) -
   2. As n --> infinity, zeta n(T) proportional to root nd.}},
DOI = {{10.1103/PhysRevLett.72.1016}},
ISSN = {{0031-9007}},
Unique-ID = {{ISI:A1994MW25600018}},
}
@article{ shraiman,
Author = {Shraiman, B. I. and Siggia, E. D.},
Title = {{Scalar turbulence}},
Journal = {{Nature}},
Year = {{2000}},
Pages = {{639-646}},
Abstract = {{The advection of a passive substance by a turbulent flow is important in
   many natural and engineering settings. The concentration of such a
   substance can exhibit complex dynamic behaviour that shows many
   phenomenological parallels with the behaviour of the turbulent velocity
   field. Yet the statistical properties of this so-called `passive scalar'
   turbulence are decoupled from those of the underlying velocity field.
   Passive scalar turbulence has recently yielded to mathematical analysis,
   and such progress may ultimately lead to a better understanding of the
   still intractable problem of fluid turbulence itself.}},
DOI = {{10.1038/35015000}},
ISSN = {{0028-0836}},
Unique-ID = {{ISI:000087465800036}},
}
@article{ ors72,
Author = {Orszag,  S. A. and Patterson, G. S.},
Title = {{Numerical simulation of 3-dimensional homogeneous isotropic turbulence}},
Journal = PRL,
Year = {{1972}},
  volume = {{28}},
Pages = {{76-\&}},
DOI = {{10.1103/PhysRevLett.28.76}},
ISSN = {{0031-9007}},
Unique-ID = {{ISI:A1972L235700003}},
}
@book{PMP,
 author = {Kirk, D. and Hwu, W. M.},
 title = {Programming Massively Parallel Processors: A Hands-on Approach},
 year = {2010},
 isbn = {0123814723, 9780123814722},
 edition = {1st},
 publisher = {Morgan Kaufmann Publishers Inc.},
 address = {San Francisco, CA, USA},
}
@book{cuda,
 author = {Sanders, J. and Kandrot, E.},
 title = {CUDA by Example: An Introduction to General-Purpose GPU Programming},
 year = {2010},
 isbn = {0131387685, 9780131387683},
 edition = {1st},
 publisher = {Addison-Wesley Professional},
} 
@article{ ishi07,
Author = {Ishihara, T. and Kaneda, Y. and Yokokawa, M. and Itakura, K. and Uno, A.},
Title = {{Small-scale statistics in high-resolution direct numerical simulation of
   turbulence: Reynolds number dependence of one-point velocity gradient
   statistics}},
Journal = JFM,
  volume = {{592}},
Year = {{2007}},
Pages = {{335-366}},
Month = {{DEC 10}},
Abstract = {{One-point statistics of velocity gradients and Eulerian and Lagrangian
   accelerations are studied by analysing the data from high-resolution
   direct numerical simulations (DNS) of turbulence in a periodic box, with
   up to 4096 3 grid points. The DNS consist of two series of runs; one is
   with k(max)eta similar to 1 (Series 1) and the other is with k(max)eta
   similar to 2 (Series 2), where k(max) is the maximum wavenumber and eta
   the Kolmogorov length scale. The maximum Taylor-microscale Reynolds
   number R-lambda in Series 1 is about 1130, and it is about 675 in Series
   2. Particular attention is paid to the possible Reynolds number (Re)
   dependence of the statistics. The visualization of the intense vorticity
   regions shows that the turbulence field at high Re consists of clusters
   of small intense vorticity regions, and their structure is to be
   distinguished from those of small eddies. The possible dependence on Re
   of the probability distribution functions of velocity gradients is
   analysed through the dependence on R-lambda of the skewness and flatness
   factors (S and F). The DNS data suggest that the R-lambda dependence of
   S and F of the longitudinal velocity gradients fit well with a simple
   power law: S similar to -0.32R(lambda)(0.11) and F similar to
   1.14(lambda)(0.34), in fairly good agreement with previous experimental
   data. They also suggest that all the fourth-order moments of velocity
   gradients scale with R-lambda similarly to each other at R-lambda > 100,
   in contrast to R-lambda < 100. Regarding the statistics of time
   derivatives, the sccond-order time derivatives of turbulent velocities
   are more intermittent than the first-order ones for both the Eulerian
   and Lagrangian velocities, and the Lagrangian time derivatives of
   turbulent velocities are more intermittent than the Eulerian time
   derivatives, as would be expected. The flatness factor of the Lagrangian
   acceleration is as large as 90 at R-lambda approximate to 430. The
   flatness factors of the Eulerian and Lagrangian accelerations increase
   with R-lambda approximately proportional to R-lambda(alpha E) and
   R-lambda(alpha L), respectively, where alpha(E) approximate to 0.5 and
   alpha(L) approximate to 1.0, while those of the second-order time
   derivatives of the Eulerian and Lagrangian velocities increases
   approximately proportional to R-lambda(beta E) and R-lambda(beta L),
   respectively, where beta(E) approximate to 1.5 and beta(L) approximate
   to 3.0.}},
DOI = {{10.1017/S0022112007008531}},
ISSN = {{0022-1120}},
Unique-ID = {{ISI:000251948800015}},
}
@article{ kaneda03,
Author = {Kaneda, Y. and Ishihara, T. and Yokokawa, M. and Itakura, K. and Uno, A.},
Title = {{Energy dissipation rate and energy spectrum in high resolution direct
   numerical simulations of turbulence in a periodic box}},
Journal = POF,
Year = {{2003}},
  volume = {{15}},
Pages = {{L21-L24}},
Month = {{FEB}},
Abstract = {{High-resolution direct numerical simulations (DNSs) of incompressible
   homogeneous turbulence in a periodic box with up to 4096(3) grid points
   were performed on the Earth Simulator computing system. DNS databases,
   including the present results, suggest that the normalized mean energy
   dissipation rate per unit mass tends to a constant, independent of the
   fluid kinematic viscosity nu as nu-->0. The DNS results also suggest
   that the energy spectrum in the inertial subrange almost follows the
   Kolmogorov k(-5/3) scaling law, where k is the wavenumber, but the
   exponent is steeper than -5/3 by about 0.1. (C) 2003 American Institute
   of Physics.}},
DOI = {{10.1063/1.1539855}},
ISSN = {{1070-6631}},
Unique-ID = {{ISI:000180316100002}},
}
@article{speziale,
    author = {Speziale, C. G.},
    journal = ARFM,
    pages = {107--157},
  volume = {23},
    title = {Analytical Methods for the Development of {Reynolds-Stress} Closures in Turbulence},
    url = {http://dx.doi.org/10.1146/annurev.fl.23.010191.000543},
    year = {1991}
}
@article{pope94,
    AUTHOR={S. B. Pope},
    journal = ARFM,
  volume = {26},
    pages = {23--63},
    title = {Lagrangian {PDF} methods for turbulent flows},
    year = {1994}
}
@article{kadoch,
author = {Kadoch, B. and Iyer, K. and Donzis, D. and Schneider, K. and Farge, M. and Yeung, P. K.},
title = {On the role of vortical structures for turbulent mixing using direct numerical simulation and wavelet-based coherent vorticity extraction},
journal = JOT,
number = {},
  volume = {{12}},
pages = {N20},
year = {2011},
doi = {10.1080/14685248.2011.562511},

URL = {http://www.tandfonline.com/doi/abs/10.1080/14685248.2011.562511},
eprint = {http://www.tandfonline.com/doi/pdf/10.1080/14685248.2011.562511}
}
@ARTICLE{SV1994,
  author = {Saddoughi, S. G. And Veeravalli, S. V.},
  title = {Local Isotropy In Turbulent Boundary-Layers At High {R}eynolds-Number},
  journal = JFM,
  year = {1994},
  volume = {268},
  pages = {333-372},
  abstract = {To test the local-isotropy predictions of Kolmogorov's (1941) universal
	equilibrium theory, we have taken hot-wire measurements of the velocity
	fluctuations in the test-section-ceiling boundary layer of the 80
	x 120 foot Full-Scale Aerodynamics Facility at NASA Ames Research
	Center, the world's largest wind tunnel. The maximum Reynolds numbers
	based on momentum thickness, R(theta), and on Taylor microscale,
	R(lambda), were approximately 370000 and 1450 respectively. These
	are the largest ever attained in laboratory boundary-layer flows.
	The boundary layer develops over a rough surface, but the Reynolds-stress
	profiles agree with canonical data sufficiently well for present
	purposes. Spectral and structure-function relations for isotropic
	turbulence were used to test the local-isotropy hypothesis, and our
	results have established the condition under which local isotropy
	can be expected. To within the accuracy of measurement, the shear-stress
	cospectral density E12(k1), which is the most sensitive indicator
	of local isotropy, fell to zero at a wavenumber about a decade larger
	than that at which the energy spectra first followed -5/3 power laws.
	At the highest Reynolds number, E12(k1) vanished about one decade
	before the start of the dissipation range, and it remained zero in
	the dissipation range. The lower wavenumber limit of locally isotropic
	behaviour of the shear-stress cospectra is given by k1(epsilon/S3)1/2
	almost-equal-to 10 where S is the mean shear, partial derivative
	U/partial derivative y. The current investigation also indicates
	that for energy spectra this limit may be relaxed to k1(epsilon/S3)1/2
	almost-equal-to 3; this is Corrsin's (1958) criterion, with the numerical
	value obtained from the present data. The existence of an isotropic
	inertial range requires that this wavenumber be much less than the
	wavenumber at the onset of viscous effects, k1 eta much less than
	1, so that the combined condition (Corrsin 1958; Uberoi 1957), is
	S(nu/epsilon)1/2 much less than 1. Among other detailed results,
	it was observed that in the dissipation range the energy spectra
	had a simple exponential decay (Kraichnan 1959) with an exponent
	prefactor close to the value 8 = 5.2 obtained in direct numerical
	simulations at low Reynolds number. The inertial-range constant for
	the three-dimensional spectrum, C, was estimated to be 1.5+/-0.1
	(Monin & Yaglom 1975). Spectral 'bumps' between the -5/3 inertial
	range and the dissipative range were observed on all the compensated
	energy spectra. The shear-stress cospectra rolled-off with a -7/3
	power law before the start of local isotropy in the energy spectra,
	and scaled linearly with S (Lumley 1967). In summary, it is shown
	that one decade of inertial subrange with truly negligible shear-stress
	cospectral density requires S(nu/epsilon)1/2 of not more than about
	0.01 (for a shear layer with turbulent kinetic energy production
	approximately equal to dissipation, a microscale Reynolds number
	of about 1500). For practical purposes many of the results of the
	hypothesis may be relied on at somewhat lower Reynolds numbers. },
  file = {:pdfs/SV1994.pdf:PDF},
  owner = {donzis},
  sn = {0022-1120},
  tc = {284},
  timestamp = {2009.04.28},
  ut = {WOS:A1994NT03800014}
}
@ARTICLE{VY.1999,
  author = {Vedula, P. and Yeung, P. K.},
  title = {Similarity scaling of acceleration and pressure statistics
        in numerical simulations of turbulence},
  journal = POF,
  volume = {{11}},
  year = {1999},
  pages = {{1208-1221}},
}
@ARTICLE{dimo05,
  author = {Dimotakis, P. E.},
  title = {Turbulent mixing},
  journal = ARFM,
  year = {2005},
  volume = {{37}},
  pages = {{329-356}},
  abstract = {{The ability of turbulent flows to effectively mix entrained fluids
	to a molecular scale is a vital part of the dynamics Of Such flows,
	with wide-ranging consequences in nature and engineering. It is a
	considerable experimental, theoretical, modeling, and computational
	challenge to capture and represent turbulent mixing which, for high
	Reynolds number (Re) flows, occurs across a spectrum of scales of
	considerable span. This consideration alone places high-Re mixing
	phenomena beyond the reach of direct simulation, especially in high
	Schmidt number fluids, such as water, in which species diffusion
	scales are one and a half orders of magnitude smaller than the smallest
	flow scales. The discussion below attempts to provide an overview
	of turbulent mixing; the attendant experimental, theoretical, and
	computational challenges; and suggests possible future directions
	for progress in this important field.}},
  file = {:dimo05.pdf:PDF},
  unique-id = {{ISI:000226822300014}}
}
@ARTICLE{bilger04,
  author = {Bilger, R. W.},
  title = {Some aspects of scalar dissipation},
  journal = {Flow, Turb. \& Combust.},
  year = {2004},
  volume = {{72}},
  pages = {{93-114}},
  abstract = {{Scalar dissipation is of great importance in the theory and modelling
	of combustion and other reacting turbulent flows. Measurements of
	scalar dissipation are found to lack the quality assurance of checks
	available from the conservation equations. Conditional averages of
	the scalar dissipation, so important in turbulent reacting flow theory
	and modelling, have qualitative and quantitative dependences that
	are very dependent on the details of the flow and mixing conditions.
	Accordingly, effort needs to focus on viable means of modelling it.
	Fluctuations of the scalar dissipation about the conditional mean
	are also important. Research results in this area need to be made
	more accessible to the combustion scientist. Heat release effects,
	so important in turbulent premixed combustion, are found to be much
	less important in non-premixed combustion.}},
  unique-id = {{ISI:000224389500002}}
}
@ARTICLE{pkzhou,
    author = {Yeung, P. K. and Zhou, Y.},
    title = {On The Universality Of The Kolmogorov Constant In Numerical Simulations Of Turbulence},
    journal = PRE,
  volume = {56},
    year = {1997},
    pages = {1746-1752}
}
@ARTICLE{DR1990,
  author = {Domaradzki, J. A. And Rogallo, R. S.},
  title = {Local energy-transfer and nonlocal interactions in homogeneous, isotropic
        turbulence},
  journal = POF,
  year = {1990},
  volume = {{2}},
  pages = {{413-426}},
  file = {:pdfs/DR1990.pdf:PDF},
  unique-id = {{ISI:A1990CR56700013}}
}
@article{ hill2002,
Author = {Hill, R},
Title = {{Structure-function equations for scalars}},
Journal = POF,
Year = {{2002}},
  volume = {{14}},
Pages = {{1745-1756}},
Month = {{May}},
DOI = {{10.1063/1.1466826}},
ISSN = {{1070-6631}},
Unique-ID = {{ISI:000174938400020}},
}
@article{ hill2006,
Author = {Hill, RJ},
Title = {{Opportunities for use of exact statistical equations}},
Journal = JOT,
Year = {{2006}},
Number = {{43}},
Pages = {{1-13}},
DOI = {{10.1080/14685240600595636}},
ISSN = {{1468-5248}},
Unique-ID = {{ISI:000238611600001}},
}
@article{Galanti,
title = "Is turbulence ergodic? ",
journal = "Phys. Lett. A ",
pages = "173 - 180",
year = "2004",
  volume = {330},
note = "",
issn = "0375-9601",
doi = "http://dx.doi.org/10.1016/j.physleta.2004.07.009",
url = "http://www.sciencedirect.com/science/article/pii/S0375960104009892",
author = "B. Galanti and A. Tsinober"
}
@Article{cooley,
  author = 	 {Cooley, J. W. and Tukey, J. W.},
  title = 	 {An algorithm for the machine calculation of complex {F}ourier series},
  journal = 	 {Math. Comput.},
  year = 	 {1965},
  pages =	 {297-301},
}
@BOOK{lesieur,
         AUTHOR={M. Lesieur},
         TITLE={Turbulence in Fluids},
         PUBLISHER={Kluwer Academic Publishers},
         edition = {3rd},
         YEAR={1997},
}
@BOOK{peyret,
         AUTHOR={R. Peyret and T. D. Taylor},
         TITLE={Computational Methods for Fluid Flow},
         PUBLISHER={Springer-Verlag},
         YEAR={1983},
}
@article{dubey,
  added-at = {2012-04-14T00:00:00.000+0200},
  author = {Dubey, A. and Tessera, D.},
  biburl = {http://www.bibsonomy.org/bibtex/2d96187adf57a189cdc93355fd7feefe3/dblp},
  ee = {http://dx.doi.org/10.1002/cpe.564},
  interhash = {4f4ca1159399f99e987c2f527be468b5},
  intrahash = {d96187adf57a189cdc93355fd7feefe3},
  journal = {Concurrency and Computation: Practice and Experience},
  keywords = {dblp},
  volume = {{13}},
  pages = {209-220},
  timestamp = {2012-04-14T00:00:00.000+0200},
  title = {Redistribution strategies for portable parallel FFT: a case study.},
  url = {http://dblp.uni-trier.de/db/journals/concurrency/concurrency13.html#DubeyT01},
  year = {2001}
}
@article{Foster,
 author = {Foster, T. and Worley, H.},
 title = {Parallel Algorithms for the Spectral Transform Method},
 journal = {SIAM J. Sci. Comput.},
 issue_date = {May 1997},
 month = may,
 year = {1997},
 issn = {1064-8275},
 pages = {806--837},
  volume = {{18}},
 numpages = {32},
 url = {http://dx.doi.org/10.1137/S1064827594266891},
 doi = {10.1137/S1064827594266891},
 acmid = {253591},
 publisher = {Society for Industrial and Applied Mathematics},
 address = {Philadelphia, PA, USA},
 keywords = {parallel algorithms, performance analysis, spectral transform method},
} 
@article{dmitry2012,
Author = {Pekurovsky, D.},
Title = {{P3DFFT: A framework for parallel computations of Fourier transforms in
   three dimensions}},
Journal = {{SIAM J. Sci. comp.}},
Year = {{2012}},
Pages = {{C192-C209}},
Abstract = {{Fourier and related transforms are a family of algorithms widely
   employed in diverse areas of computational science, notoriously
   difficult to scale on high-performance parallel computers with a large
   number of processing elements (cores). This paper introduces a popular
   software package called P3DFFT which implements fast Fourier transforms
   (FFTs) in three dimensions in a highly efficient and scalable way. It
   overcomes a well-known scalability bottleneck of three-dimensional (3D)
   FFT implementations by using two-dimensional domain decomposition.
   Designed for portable performance, P3DFFT achieves excellent timings for
   a number of systems and problem sizes. On a Cray XT5 system P3DFFT
   attains 45\% efficiency in weak scaling from 128 to 65,536 computational
   cores. Library features include Fourier and Chebyshev transforms,
   Fortran and C interfaces, in- and out-of-place transforms, uneven data
   grids, and single and double precision. P3DFFT is available as open
   source at http://code.google.com/p/p3dfft/. This paper discusses P3DFFT
   implementation and performance in a way that helps guide the user in
   making optimal choices for parameters of their runs.}},
DOI = {{10.1137/11082748X}},
ISSN = {{1064-8275}},
Unique-ID = {{ISI:000310475700033}},
}
@InProceedings{bgfft,
 author 	= 	{A. Chan and P. Balaji and W. Gropp and R. Thakur},
	title 	= 	{Communication Analysis of Parallel {3D FFT} for Flat {C}artesian Meshes on Large {Blue Gene} Systems},
	booktitle 	= 	{15th IEEE Int. Conf. on High Perf. Comp.},
	year 	= 	2008,
	pages 	= 	{422-429},
	area 	= 	"App:Par:M:Coll",
} 
@article{elecstaturb,
Author = {F. {L}epreti and V. Carbone and M. Spolaore and V. Antoni and
   R. Cavazzana and E. Martines and G. Serianni and P. Veltri and
   N. Vianello and M. Zuin},
Title = {{Yaglom law for electrostatic turbulence in laboratory magnetized plasmas}},
Journal = EPL,
Year = {{2009}},
pages =	{1-5},
Volume = {{86}},
Month = {{APR}},
Abstract = {{It has recently been shown that a Yaglom law for electrostatic
   turbulence, that is, a relation for the third-order mixed moment
   involving the particle number density as a passive scalar and the E x B
   drift velocity, can be deduced from a simple model of electrostatic
   fluctuations, which describes bursty turbulence in plasmas. In this
   letter, the existence of the Yaglom law for electrostatic turbulence in
   laboratory magnetized plasmas is reported for the first time. Using
   measurements of intermittent transport at the edge of the RFX-mod
   reversed field pinch plasma device, we found that the above scaling
   relation is nicely verified at intermediate scales of few centimeters.
   In this range of scales, that unambiguously represents the inertial
   range of electrostatic turbulence, we also analyze the intermittency
   properties of electrostatic turbulence by measuring anomalous scaling
   exponents of density and velocity structure functions. Copyright (C)
   EPLA, 2009}},
DOI = {{10.1209/0295-5075/86/25001}},
Article-Number = {{25001}},
ISSN = {{0295-5075}},
ResearcherID-Numbers = {{Martines, Emilio/B-1418-2009
   Vianello, Nicola/B-6323-2008}},
ORCID-Numbers = {{Martines, Emilio/0000-0002-4181-2959
   Vianello, Nicola/0000-0003-4401-5346}},
Unique-ID = {{ISI:000266414100019}},
}

@article{ mhdturb,
Author = {Sorriso-Valvo, L. and Marino, R. and Carbone, V. and Noullez, A. and
   Lepreti, F. and Veltri, P. and Bruno, R. and Bavassano, B. and
   Pietropaolo, E.},
Title = {{Observation of inertial energy cascade in interplanetary space plasma}},
Journal = PRL,
Year = {{2007}},
Volume = {{99}},
Number = {{11}},
Month = {{SEP 14}},
Abstract = {{Direct evidence for the presence of an inertial energy cascade, the most
   characteristic signature of hydromagnetic turbulence (MHD), is observed
   in the solar wind by the Ulysses spacecraft. After a brief rederivation
   of the equivalent of Yaglom's law for MHD turbulence, a linear relation
   is indeed observed for the scaling of mixed third-order structure
   functions involving Elsasser variables. This experimental result firmly
   establishes the turbulent character of low-frequency velocity and
   magnetic field fluctuations in the solar wind plasma.}},
DOI = {{10.1103/PhysRevLett.99.115001}},
Article-Number = {{115001}},
ISSN = {{0031-9007}},
ResearcherID-Numbers = {{sorriso-valvo, luca/A-9355-2008}},
Unique-ID = {{ISI:000249474700032}},
}
@ARTICLE{YDS2005,
  author = {Yeung, P. K. and Donzis, D. A. and Sreenivasan, K. R.},
  title = {{High-Reynolds-number simulation of turbulent mixing}},
  journal = POF,
  year = {2005},
  volume = {{17}},
  pages = {081703},
  abstract = {{A brief report is given of a new 2048(3) direct numerical simulation
	of the mixing of passive scalars with uniform mean gradients in forced,
	stationary isotropic turbulence. The Taylor-scale Reynolds number
	is close to 700 and Schmidt numbers of 1 and 1/8 are considered.
	The data provide the most convincing evidence to date for the inertial-convective
	scaling. Significant departures from small-scale isotropy are sustained
	in conventional measures. Subject to some stringent resolution requirements,
	the data suggest that commonly observed differences between the intermittency
	of energy and scalar dissipation rates may in part be a finite-Reynolds-number
	effect. (c) 2005 American Institute of Physics.}},
  article-number = {{081703}},
  owner = {donzis},
  timestamp = {2008.07.08},
  unique-id = {{ISI:000231392000003}}
}
@ARTICLE{sreeni1996,
  author = {Sreenivasan, K. R.},
  title = {The passive scalar spectrum and the Obukhov-Corrsin constant},
  journal = {Phys. Fluids},
  year = {1996},
  volume = {8},
  pages = {189--196},
  abstract = {It is pointed out that, for microscale Reynolds numbers less than
	about 1000, the passive scalar spectrum in turbulent shear flows
	is less steep than anticipated and that the Obukhov-Corrsin constant
	can be defined only if the microscale Reynolds number exceeds this
	value. In flows where the large-scale velocity field is essentially
	isotropic (as in grid turbulence), the expected 5/3 scaling is observed
	even at modest Reynolds numbers. All known data on the Obukhov-Corrsin
	constant are collected. The support for the notion of a ''universal''
	constant is shown to be reasonable. Its value is about 0.4. (C) 1996
	American Institute of Physics. },
  file = {:pdfs/sreeni1996.pdf:PDF},
  owner = {donzis},
  sn = {1070-6631},
  tc = {91},
  timestamp = {2009.12.02},
  ut = {WOS:A1996TL89200019}
}
@ARTICLE{krs95,
  author = {Sreenivasan, K. R.},
  title = {On the universality of the Kolmogorov constant},
  journal = POF,
  year = {1995},
  volume = {7},
  pages = {2778--2784},
}
@ARTICLE{Pumir.1994,
  author = {Pumir, A.},
  title = {A numerical study of pressure fluctuations in three-dimensional, incompressible, homogeneous, tropic turbulence},
  journal = POF,
  year = {1994},
  volume = {{6}},
  pages = {{2071-2083}},
}
@article{Pumir94,
 author               = {Pumir, A.},
 journal              = POF,
 pages                = {2118--2132},
 title                = {A numerical study of the mixing of a passive scalar in three dimensions in the presence of a mean gradient},
 volume               = {6},
 year                 = {1994},
 }
@article{TKE03,
  title = {Recovering isotropic statistics in turbulence simulations: The Kolmogorov 4/5th law},
  author = {Taylor, M. A. and Kurien, S. and Eyink, G. L.},
  journal = PRE,
  issue = {2},
  pages = {026310},
  numpages = {8},
  year = {2003},
  month = {Aug},
  doi = {10.1103/PhysRevE.68.026310},
  url = {http://link.aps.org/doi/10.1103/PhysRevE.68.026310},
  publisher = {American Physical Society}
}
@article{eyink02,
Author = {Eyink, G},
Title = {{Local 4/5-law and energy dissipation anomaly in turbulence}},
Journal = {{{N}onlinearity}},
Year = {{2003}},
Volume = {{16}},
Pages = {{137-145}},
Month = {{JAN}},
Abstract = {{A strong local form of the `4/3-law' in turbulent flow has been proved
   recently by Duchon and Robert for a triple moment of velocity increments
   averaged over both a bounded spacetime region and separation vector
   directions, and for energy dissipation averaged over the same spacetime
   region. Under precisely stated hypotheses, the two are proved to be
   proportional, by a constant 4/3, and to appear as a non-negative defect
   measure in the local energy balance of singular (distributional)
   solutions of the incompressible Euler equations. Here, we prove that the
   energy defect measure can also be represented by a triple moment of
   purely longitudinal velocity increments and by a mixed moment with one
   longitudinal and two tranverse velocity increments. Thus, we prove that
   the traditional 4/5- and 4/15-laws of Kolmogorov hold in the same local
   sense as demonstrated for the 4/3-law by Duchon-Robert.}},
DOI = {{10.1088/0951-7715/16/1/309}},
ISSN = {{0951-7715}},
Unique-ID = {{ISI:000180614500010}},
}
@article{ nie99,
Author = {Nie, Q and Tanveer, S},
Title = {{A note on third-order structure functions in turbulence}},
Journal = {{Proc. Royal Soc. A-Mathematical  physical and Engineering
   Sciences}},
Year = {{1999}},
Volume = {{455}},
Pages = {{1615-1635}},
Month = {{MAY 8}},
ISSN = {{1364-5021}},
EISSN = {{1471-2946}},
Unique-ID = {{ISI:000080400600002}},
}
@ARTICLE{GFN2002,
  author = {Gotoh, G. and Fukayama, D. and Nakano, T.},
  title = {Velocity field statistics in homogeneous steady turbulence obtained
	using a high-resolution direct numerical simulation},
  journal = POF,
  year = {2002},
  volume = {14},
  pages = {1065-1081},
  file = {:pdfs/GFN2002.pdf:PDF},
  keywords = {numerical analysis; flow simulation; turbulence; Navier-Stokes equations},
  publisher = {AIP},
  url = {http://link.aip.org/link/?PHF/14/1065/1}
}
@article{ antonia2006,
Author = {Antonia, R. A. and Burattini, P},
Title = {{Approach to the 4/5 law in homogeneous isotropic turbulence}},
Journal = JFM,
Year = {{2006}},
Volume = {{550}},
Pages = {{175-184}},
Month = {{MAR 10}},
Abstract = {{Kolmogorov's similarity hypotheses and his 4/5 law are valid at very
   large Reynolds numbers. For flows encountered in the laboratory, the
   effect of a finite Reynolds number and of the non-stationarity or
   inhomogeneity associated with the large scales can affect the behaviour
   of the scales in the inertial range significantly. This paper focuses on
   the source of inhomogeneity in two types of flows, those dominated
   mainly by a decay of energy in the streamwise direction and those which
   are forced, through a continuous injection of energy at large scales.
   Results based on a parameterization of the second-order velocity
   structure function indicate that the normalized third-order structure
   function approaches 4/5 much more rapidly for forced than for decaying
   turbulence. This trend is supported by grid turbulence measurements and
   numerical data in a periodic box.}},
DOI = {{10.1017/S0022112005008438}},
ISSN = {{0022-1120}},
Unique-ID = {{ISI:000236195700011}},
}
@BOOK{MY.I,
  title={Statistical Fluid Mechanics},
  publisher={MIT Press},
  year={1975},
  author={Monin, A. S. and Yaglom, A. M.},
  volume={1}
}
@BOOK{MY.II,
  title={Statistical Fluid Mechanics},
  publisher={MIT Press},
  year={1975},
  author={Monin, A. S. and Yaglom, A. M.},
  volume={2}
}
@article{ duchon02,
Author = {Duchon, J and Robert, R},
Title = {{Inertial energy dissipation for weak solutions of incompressible Euler
   and Navier-Stokes equations}},
Journal = {{Nonlinearity}},
Year = {{2000}},
Volume = {{13}},
Pages = {{249-255}},
Month = {{JAN}},
Abstract = {{We study the local equation of energy for weak solutions of
   three-dimensional incompressible Navier-Stokes and Euler equations. We
   define a dissipation term D(u) which stems from an eventual lack of
   smoothness in the solution u. We give in passing a simple proof of
   Onsager's conjecture on energy conservation for the three-dimensional
   Euler equation, slightly weakening the assumption of Constantin et al.
   We suggest calling weak solutions with non-negative D(u) `dissipative'.}},
DOI = {{10.1088/0951-7715/13/1/312}},
ISSN = {{0951-7715}},
Unique-ID = {{ISI:000084894300013}},
}
@article{renka1,
Author = {Renka, R},
Title = {{Interpolation of data on the surface of a sphere}},
 journal = {ACM Trans. Math. Softw.},
Year = {{1984}},
Volume = {{10}},
Pages = {{417-436}},
DOI = {{10.1145/2701.2703}},
ISSN = {{0098-3500}},
Unique-ID = {{ISI:A1984ABJ1400007}},
}
@article{renka2,
Author = {Renka, RJ},
Title = {{Interpolation on the surface of a sphere}},
Journal = {{ACM Trans. ON Math. software}},
Year = {{1984}},
Volume = {{10}},
Pages = {{437-439}},
DOI = {{10.1145/2701.356107}},
ISSN = {{0098-3500}},
Unique-ID = {{ISI:A1984ABJ1400008}},
}
@book{okabe,
 author = {Okabe, A and Boots, B and Sugihara, K},
 title = {Spatial Tessellations: Concepts and Applications of Voronoi Diagrams},
 year = {1992},
 isbn = {0-471-93430-5},
 publisher = {John Wiley \&amp; Sons, Inc.},
 address = {New York, NY, USA},
}
@article{Renka97,
 author = {Renka, R.},
 title = {Algorithm 772: STRIPACK: Delaunay Triangulation and Voronoi Diagram on the Surface of a Sphere},
 journal = {ACM Trans. Math. Softw.},
 issue_date = {Sept. 1997},
 volume = {23},
 month = sep,
 year = {1997},
 issn = {0098-3500},
 pages = {416--434},
 numpages = {19},
 url = {http://doi.acm.org/10.1145/275323.275329},
 doi = {10.1145/275323.275329},
 acmid = {275329},
 publisher = {ACM},
 address = {New York, NY, USA},
 keywords = {Delaunay triangulation, Dirichlet tessellation, Thiessen regions, Voronoi diagram, sphere},
}
@techreport{blelloch,
	author = {Blelloch, G.},
	title = {Prefix Sums and Their Applications},
	institution = {School of Computer Science, Carnegie Mellon University},
	number = {CMU-CS-90-190},
	month = nov,
	year = {1990}
}
@article{ biff02,
Author = {Biferale, L and Lohse, D and Mazzitelli, I and Toschi, F},
Title = {{Probing structures in channel flow through SO(3) and SO(2) decomposition}},
Journal = JFM,
Year = {{2002}},
Volume = {{452}},
Pages = {{39-59}},
Month = {{FEB 10}},
Abstract = {{SO(3) and SO(2) decompositions of numerical channel flow turbulence are
   performed. The decompositions are used to probe, characterize, and
   quantify anisotropic structures in the flow. Close to the wall, the
   anisotropic modes are dominant and reveal the flow structures. The
   dominance of the (j, m) = (2,1) mode of the SO(3) decomposition in the
   buffer layer is associated with hairpin vortices. The SO(2)
   decomposition in planes parallel to the walls allows us also to access
   the regions very close to the wall. In those regions we have found that
   the strong enhancement of intermittency can be explained in terms of
   streaklike structures and their signatures in the in = 2 and m = 4 modes
   of the SO(2) decomposition.}},
DOI = {{10.1017/S0022112001006632}},
ISSN = {{0022-1120}},
ResearcherID-Numbers = {{Toschi, Federico/A-1552-2009
   Lohse, Detlef/B-4915-2013
   biferale, luca/L-4535-2013}},
ORCID-Numbers = {{Toschi, Federico/0000-0001-5935-2332
   Lohse, Detlef/0000-0003-4138-2255
   }},
Unique-ID = {{ISI:000174197200003}},
}
@article{ gotoh12,
Author = {Gotoh, T. and Hatanaka, S. and Miura, H.},
Title = {{Spectral compact difference hybrid computation of passive scalar in
   isotropic turbulence}},
Journal = {{J. {C}omp. {P}hys.}},
Year = {{2012}},
Volume = {{231}},
Pages = {{7398-7414}},
Month = {{AUG 30}},
Abstract = {{An accurate and efficient hybrid numerical method is developed for
   direct numerical simulation of passive scalar in homogeneous turbulence
   for the Schmidt number 1 and 50. The hybrid method uses the standard
   Fourier spectral method for the incompressible Navier-Stokes equation
   and the combined compact difference scheme for the passive scalar
   transport equation. Accuracy of the method is carefully examined by
   comparing with the full spectral method regarding the spectra,
   probability density function, field structure of the passive scalar, and
   is found to be very satisfactory. The computational time for the hybrid
   method is decreased by 26\% for the Schmidt number 1 when compared to
   the full spectral method, and by 77\% for the Schmidt number 50 when the
   number of grid points for the velocity field is reduced under the scale
   separation. (C) 2012 Elsevier Inc. All rights reserved.}},
DOI = {{10.1016/j.jcp.2012.07.010}},
ISSN = {{0021-9991}},
Unique-ID = {{ISI:000308067000020}},
}
@inproceedings{vuducfft,
  acceptance = {[36/161=22.4
  address = {San Servolo Island, Venice, Italy},
  author = {Czechowski, T. and McClanahan, C. and Battaglino, C. and K. P. Iyer and P. K. Yeung and Vuduc R.},
  booktitle = {Proc.~ACM Int'l. Conf. Supercomputing (ICS)},
  date-added = {2012-03-15 15:12:43 +0000},
  date-modified = {2014-02-14 16:49:32 +0000},
  doi = {10.1145/2304576.2304604},
  month = {June},
  pdf = {czechowski2012-ics-xfft.pdf},
  talk = {czechowski2012-ics-xfft--SLIDES.pdf},
  title = {On the communication complexity of {3D} {FFTs} and its implications for exascale},
  topic = {FFT; exascale; performance modeling; co-design},
  year = {2012},
  bdsk-url-1 = {http://dx.doi.org/10.1145/2304576.2304604}
}
@ARTICLE{KPPK14,
  AUTHOR={K. P. Iyer and P. K. Yeung},
  TITLE={Structure functions and applicability of {Y}aglom's relation in
passive-scalar turbulent mixing at low Schmidt numbers with uniform
mean gradient},
  JOURNAL=POF,
 volume = {submitted},
  YEAR={2014}}
@PHDTHESIS{kp2014,
   author       = "K. P. Iyer",
   title        = "Studies of turbulence structure and turbulent mixing using Petascale computing",
   school       = "Georgia Institute of  Technology",
   year         = "2014",

   type         = "{Ph.D.} thesis",
   address      = "",
   month        = "",
   note         = "",
}
@article{obukhov62,
Author = {{O}bukhov, A. M.},
Title = {{S}ome specific features of atmospheric turbulence},
Journal = JFM,
volume = {13},
Year = {1962},
Pages = {77-81},
DOI = {{10.1017/S0022112062000506}},
ISSN = {{0022-1120}},
Unique-ID = {{ISI:A1962WT24400006}},
}
@article{goto14,
Author = {{S. Goto, A. Matsunaga, M. Fujiwara, M. Nishioka,
         S. Kida, M. Yamato, M. and S. Tsuda}},
Title = {{Turbulence driven by precession in spherical and slightly elongated spheroidal cavities}},
Journal = POF,
volume = {26},
Year = {2014},
Pages = {{055107}}
}
@article{kida11,
author = {{S. Kida}},
title = {{Steady flow in a rapidly rotating sphere with weak precession}},
journal = JFM,
volume = {680},
year = {2011},
pages = {150--193}
}
@article{goto07,
author = {{S. Goto, N. Ishii, S. Kida, and M. Nishioka}},
Journal = POF,
Pages = {061705},
Title = {{Turbulence generator using a precessing sphere}},
Volume = {19},
Year = {2007}}
@article{meunier,
author = {{Meunier, P. and Eloy, C. and Lagrange, R. and Nadal, F.}},
title = {A rotating fluid cylinder subject to weak precession},
journal = JFM,
volume = {599},
month = {3},
year = {2008},
issn = {1469-7645},
pages = {405--440},
numpages = {36},
}
@article{Malkus,
author = {{W. V. R. Malkus}}, 
title = {{Precession of the Earth as the Cause of Geomagnetism: Experiments lend support to the proposal that precessional torques drive the earth's dynamo}},
volume = {160}, 
pages = {259-264}, 
year = {1968}, 
abstract ={I have proposed that the precessional torques acting on the earth can sustain a turbulent hydromagnetic flow in the molten core. A gross balance of the Coriolis force, the Lorentz force, and the precessional force in the core fluid provided estimates of the fluid velocity and the interior magnetic field characteristic of such flow. Then these numbers and a balance of the processes responsible for the decay and regeneration of the magnetic field provided an estimate of the magnetic field external to the core. This external field is in keeping with the observations, but its value is dependent upon the speculative value for the electrical conductivity of core material. The proposal that turbulent  flow due to precession can occur in the core was tested in a study of nonmagnetic laboratory flows induced by the steady precession of fluid-filled rotating spheroids. It was found that these flows exhibit both small wavelike instabilities and violent finite-amplitude instability to turbulent motion above critical values of the precession rate. The observed critical parameters indicate that a laminar flow in the core, due to the earth's precession, would have weak hydrodynamic instabilities at most, but that finite-amplitude hydromagnetic instability could lead to fully turbulent flow.}, 
journal = {{Science}} 
}
@article{camb97,
author = {{C. Cambon, N. N. Mansour and F. S. Godeferd}},
title = {{Energy transfer in rotating turbulence}},
journal = JFM,
volume = {227},
year = {1997},
pages = {303--332}
}
@article{map09,
   author = "Mininni, P. D. and Alexakis, A. and Pouquet, A.",
   title = "Scale interactions and scaling laws in rotating flows at moderate Rossby numbers and large Reynolds numbers",
   journal = POF,
   year = "2009",
   volume = "21",
   number = "1", 
   eid = 015108,
   pages = "-",
}
@ARTICLE{pkzhou98,
    author = {{Yeung, P. K. and Zhou, Y.}},
    title = {{Numerical study of rotating turbulence with external forcing}},
    journal = POF,
  volume = {10},
    year = {1998},
    pages = {2895-2909}
}
@article {Pouquet2010,
	author = {{A. Pouquet and P. D. Mininni}},
	title = {{The interplay between helicity and rotation in turbulence: implications for scaling laws and small-scale dynamics}},
	volume = {368},
	pages = {1635--1662},
	year = {2010},
	journal = PTRSA
}
@article{sen12,
  title = {{Anisotropy and nonuniversality in scaling laws of the large-scale energy spectrum in rotating turbulence}},
  author = {{A. Sen, P. D. Mininni, D. Rosenberg and A. Pouquet}},
  journal = PRE,
  volume = {86},
  pages = {036319},
  year = {2012},
}
@article{min12,
  title = {{Isotropization at small scales of rotating
helically driven turbulence}},
  author = {{P. D. Mininni, D. Rosenberg and A. Pouquet}},
  journal = JFM,
  volume = {699},
  pages = {263-279},
  year = {2012}
}
@article{cher07,
Author = {{M. Chertkov, C. Connaughton, I. Kolokolov and V. Lebedev}},
Title = {{Dynamics of Energy Condensation in Two-Dimensional Turbulence}},
Journal = PRL,
Year = {2007},
volume = {99},
Pages = {084501}
}
@article{SY93,
Author = {{L. M. Smith and V. Yakhot}},
Title = {{Bose Condensation and Small-Scale Structure Generation in a Random Force Driven 2D Turbulence}},
Journal = PRL,
Year = {1993},
volume = {71},
Pages = {352-355}
}
@article{borue,
Author = {{Borue V.}},
Title = {{Inverse energy cascade in stationary two-dimensional homogeneous turbulence}},
Journal = PRL,
Year = {1994},
volume = {72},
Pages = {1475}
}
@article{xia11,
Author = {{H. Xia, D. Byrne, G. Falkovich and M. Shats}},
Title = {Upscale energy transfer in thick turbulent fluid layers},
Journal = {Nature Phys.},
Year = {2011},
volume = {7},
Pages = {321-324}
}
@article{BMT12,
Author = {{Biferale, L. and Musacchio, S. and Toschi F.}},
Title = {{Inverse Energy Cascade in Three-Dimensional Isotropic Turbulence}},
Journal = PRL,
Year = {2012},
volume = {108},
Pages = {164501}
}
@BOOK{gspan,
  AUTHOR={{H. P. Greenspan}},
  TITLE={{The Theory of Rotating Fluids}},
  YEAR={1990},
  PUBLISHER={{Cambridge Monographs on Mech. \& App. Math., Breukelen Press}}
}
@BOOK{gspan1,
  AUTHOR={{H. P. Greenspan}},
  TITLE={{The Theory of Rotating Fluids}},
  YEAR={1990},
  PUBLISHER={Cambridge Monographs on Mechanics and Applied Mathematics, Breukelen Press}
}
@article{saw91,
   author = {{B. L. Sawford}},
   title = {{Reynolds number effects in Lagrangian stochastic models of turbulent dispersion}},
   journal = POF ,
   year = {1991},
   volume = {3},
   pages = {1577-1586}
}
@article{zee94,
   author = {{O. Zeman}},
   title = {{A note on the spectra and decay of rotating homogeneous turbulence}},
   journal = POF,
   year = {1994},
   volume = {6},
   pages = {3221-3223}
}
@article{zhou95,
   author = {{Y. Zhou}},
   title = {{A phenomenological treatment of rotating turbulence}},
   journal = POF,
   year = {1995},
   volume = {7},
   pages = {2092-2094}
}
@article{mans91,
   author = {{N. N. Mansour, T. Shih and W. C. Reynolds}},
   title = {{The effects of rotation on initially anisotropic homogeneous flows}},
   journal = POF,
   year = {1991},
   volume = {3},
   pages = {2421-2425}
}
@article{SM14,
   author = {{B. Saint-Michel, B. Dubrulle, L. Marié, F. Ravelet and F. Daviaud}},
   title = {{Influence of Reynolds number and forcing type in a
turbulent von K\'{a}rm\'{a}n flow}},
   journal = NJP,
   year = {2014},
   volume = {16},
   pages = {063037}
}
@article{SM15,
   author = {{S. Thalabard, B. Saint-Michel, E. Herbert,
F. Daviaud and B. Dubrulle}},
   title = {{A statistical mechanics framework for the large-scale 
structure of turbulent von K\'{a}rm\'{a}n flows}},
   journal = NJP,
   year = {2015},
   volume = {17},
   pages = {063006}
}
@article{nore,
  author={{C. Nore, J. L\'eorat, J.-L. Guermond and F. Luddens}},
  title={{Nonlinear dynamo action in a cylindrical container driven by precession}},
  journal={{J. Phys.: Conf. Series}},
  volume={318},
  pages={072034},
  year={2011}
}
@article{triana,
  author={{S. A. Triana, D. S. Zimmerman and D. P. Lathrop}},
  title={{Precessional states in a laboratory model of the Earth’s core}},
  journal={{J. Geophys. Res.}},
  volume={117},
  pages={B04103},
  year={2012}
}
@article{fabien,
  author={{F. S. Godeferd and F. Moisy}},
  title={{Structure and Dynamics of Rotating Turbulence: A Review of Recent Experimental and Numerical Results}},
  journal={{App. Mech. Rev.}},
  volume={67},
  pages={030802},
  year={2015}
}
@article{LP05,
  author={{L. Biferale and I. Procaccia}},
  title={{Anisotropy in turbulent flows and in turbulent transport}},
  journal={{Phys. Rep.}},
  volume={414},
  pages={43-164},
  year={2005}
}
@article{kai05,
  author={{K. Schneider}},
  title={{Numerical simulation of the transient flow behaviour in
chemical reactors using a penalisation method}},
  journal={{Computers \& Fluids}},
  volume={34},
  pages={1223-1238},
  year={2005}
}

\end{document}